\documentclass[
  aps,pra,twocolumn,showkeys,noprintnumbers,
  amssymb,superscriptaddress,longbibliography
]{revtex4-2}

\usepackage{graphicx}
\usepackage{bm}
\usepackage{amsmath,amssymb}
\usepackage{mathtools}
\usepackage{braket}
\usepackage{physics}
\usepackage{dsfont}
\usepackage{gensymb}
\usepackage{latexsym}

\usepackage[
  breaklinks,
  colorlinks,
  citecolor=blue,
  linkcolor=red,
  urlcolor=blue
]{hyperref}

\usepackage{float}
\makeatletter
\let\newfloat\newfloat@ltx
\makeatother

\usepackage{algorithm}
\usepackage{algpseudocode}

\usepackage{cleveref}
\crefname{equation}{Eq.}{Eqs.}
\Crefname{equation}{Eq.}{Eqs.}
\crefname{figure}{Fig.}{Figs.}
\Crefname{figure}{Fig.}{Figs.}
\crefname{section}{Sec.}{Secs.}
\Crefname{section}{Sec.}{Secs.}
\crefname{algorithm}{Alg.}{Algs.}
\Crefname{algorithm}{Alg.}{Algs.}

\usepackage{xcolor}
\newcommand{\change}[1]{#1}

\usepackage{appendix}
\usepackage{amsthm}

\usepackage{listings}

\usepackage{minted}

\begin{document}

\title{Continuous Noise Model for Quantum Circuits}

\author{Yunos El Kaderi}
\affiliation{Laboratoire de Physique Th\'eorique et
Mod\'elisation (LPTM), CNRS UMR 8089, CY Cergy Paris Universit\'e, Cergy-Pontoise, France}\affiliation{\'Equipe Traitement de l’Information et Syst\`emes (ETIS), CNRS UMR 8051, CY Cergy Paris Universit\'e, Cergy-Pontoise, France}

\author{Andreas Honecker}
\affiliation{Laboratoire de Physique Th\'eorique et
Mod\'elisation (LPTM), CNRS UMR 8089, CY Cergy Paris Universit\'e, Cergy-Pontoise, France}

\author{Iryna Andriyanova}
\affiliation{\'Equipe Traitement de l’Information et Syst\`emes (ETIS), CNRS UMR 8051, CY Cergy Paris Universit\'e, Cergy-Pontoise, France}

\date{\today}

\begin{abstract}
Quantum noise is a central challenge in quantum computing across many applications. Extensive work has examined how qubits couple to their environment, leading to decoherence and relaxation, which is irreversible. Current studies focus on coherent gate errors caused by control misalignment, which accumulate with circuit depth but can, in principle, be corrected. This work studies a continuous coherent noise model for quantum circuits and compares it with a discrete Pauli model. The focus is on small coherent gate errors that build up across circuit depth. These errors are modeled as random rotations on the Bloch sphere using a von Mises--Fisher distribution. In the small-angle limit, the model reduces to an isotropic Gaussian distribution. We test the model on quantum error-correction circuits based on the $[[5,1,3]]$ and $[[7,1,3]]$ codes. A variant of Grover’s search circuit with different qubit counts is also examined. To enable fair comparison, we introduce a model-independent matching scheme. Pauli and continuous noise channels are aligned using the binary entropy at readout. This isolates the effect of noise structure at fixed uncertainty. An approximate analytical method for the propagation of coherent errors is also developed. The method tracks error distributions both on Clifford and non-Clifford circuits without full Monte Carlo sampling. It reduces simulation cost while preserving accuracy for circuit-level error estimates. The approximation is validated against brute-force simulations, identifying its regime of validity with Clifford and non-Clifford circuits
under error correction.
Our results show that continuous coherent noise and Pauli noise lead to comparable logical-error trends in the stabilizer-code circuits studied here. This behavior is consistent with syndrome extraction projecting a general single-qubit error onto Pauli-error subspaces before recovery. They also show where simplified propagation models work well and where their accuracy is reduced.
\end{abstract}

\newcommand{\qubit}{\text{qubit}}
\newcommand{\shots}{N_{\text{shots}}}
\newcommand{\numcircuits}{N_{\text{circuits}}}
\newcommand{\numgates}{N_{\text{gates}}}
\newcommand{\rad}{\text{ rad}}

\maketitle

\section{Introduction}

Quantum computing promises clear advantages for tasks such as integer factorization, quantum simulation, and certain optimization problems \cite{Montanaro2016,Bharti2022}. In practice, these gains are limited by noise \cite{Cai2023}. Decoherence, gate errors, and readout faults reduce fidelity and restrict circuit depth \cite{Cai2023,TERA2026132127}. This issue becomes more severe as the system size increases \cite{Bharti2022}. Larger quantum processors require more gates and longer runtimes, which amplify errors. As a result, accurate noise models and effective mitigation methods are required for reliable computation and for the development of fault-tolerant error correction \cite{Cai2023,Terhal2015,Tripathi2025}.

A central difficulty is that physical noise is not simple. Standard models describe errors as stochastic Pauli faults, such as bit flips and phase flips \cite{Terhal2015}.
Current quantum error correction approaches rely heavily on this picture \cite{Lescanne2020,Acharya2023,Bravyi2024,Acharya2025,Ruiz2025,rousseau2025enhancingdissipativecatqubit}. This is often an effective description, but it may not describe all gate-level effects present in real devices. An important advantage of such Pauli-based models, in particular the depolarizing channel, is that they are compatible with efficient classical Monte Carlo simulation via the Gottesman–Knill theorem: stabilizer circuits involving only Clifford operations can be simulated in polynomial time on a classical computer \cite{gottesman1998heisenbergrepresentationquantumcomputers,Aaronson2004,stim}. However, this efficiency comes at a cost, as it restricts the dynamics to Clifford operations, therefore excluding non-Clifford effects. In hardware, gate errors often arise from control drift, detuning, and coupling to uncontrolled environmental modes \cite{TERA2026132127,Tripathi2025}. These effects generate coherent errors that shift unitary operations. They accumulate over time and across layers. \change{This gives rise to dynamics that is continuous and structured rather than purely discrete.}

\change{The continuous noise model represents small gate errors as random unitary shifts on the Bloch sphere.} Recent studies show that coherent errors often follow directional patterns. This supports the use of distributions such as the von~Mises--Fisher (vMF) law~\cite{Fisher1953} to model the spread of rotation angles. Experiments with superconducting qubits~\cite{Sheldon2016,Kaufmann_2025} show clear angular biases rather than random Pauli events, which points to smooth and directional noise. Theory work, including spin-Wigner tools~\cite{Hanks2024}, shows that these unitary shifts can be captured with low-parameter continuous models that enable new mitigation ideas.

Recent theory links the vMF law to continuous noise in states and gates. Schultz~\cite{Schultz2016} showed that vMF and matrix-Fisher models provide
a natural way to describe random unitaries on the Bloch sphere. Wudarski \emph{et~al.}~\cite{Wudarski2020} used vMF-distributed states to show how anisotropic noise affects fidelity in ways that Haar sampling hides. Peng \emph{et~al.}~\cite{Peng2025} applied vMF and Watson models to free-space polarization qubits and tied the vMF concentration $\kappa$ to the strength of depolarization. Similar ideas appear in metrology, where vMF noise shapes phase-estimation limits~\cite{Maiti2022}, and in variational circuits, where vMF priors model smooth coherent drift in prepared states~\cite{Huynh2024}. Together, these works show that the vMF family gives a clear and physical way to treat coherent, Gaussian-like noise.

\change{Evidence from experiments and theory motivates noise models that go beyond purely stochastic Pauli faults.} Work on superconducting qubits~\cite{Sheldon2016} shows that repeated gate applications can separate coherent unitary errors from non-unitary noise. This is relevant for small over- or under-rotations, since such errors can build up coherently across a circuit. Other studies~\cite{Thomas2011,GeneralizedPhaseEstimation} treat rotation-angle and rotation-axis errors with geometric tools and continuous angular distributions, including von~Mises--Fisher noise. These works support the use of directional noise models for gate-level errors.

Large-scale simulations~\cite{Barnes2017,Huang2019} also show that coherent errors can behave differently from stochastic Pauli errors after error correction. In particular, coherent noise can produce broad failure distributions and logical channels that are not captured by a simple one-to-one map to depolarizing noise. At the hardware level, leakage and bosonic encodings provide further examples where non-Pauli structure matters~\cite{Marshall2025,Noh2022}. Leakage can persist outside the computational subspace, while GKP codes naturally involve continuous displacement errors and analog syndrome information. These results support using continuous or coherent noise models as a complement to Pauli-based descriptions.

In parallel, surface-code studies have shown that coherent errors can change logical-level behavior in ways that simple Pauli approximations may miss~\cite{Bravyi2018,Marton2023coherenterrors}.
\change{These works show that coherent noise can produce structured logical faults, even when the effective logical noise becomes more stochastic after error correction.} They also motivate mitigation strategies such as twirling, randomized compiling, or noise-aware decoding, depending on the hardware and circuit setting. This supports the use of realistic, directional noise models like the one studied here.

This paper studies a continuous coherent noise model and compares it with a discrete Pauli model. We focus on small, coherent gate errors that accumulate across quantum circuits.\\
We test a Gaussian small-angle limit of a von Mises--Fisher rotational model on key circuits, introduced in Sec.~\ref{sec:continuous_errors}. The test case circuits are defined in Sec.~\ref{QuantumCircuitsStudied}, which includes coded and uncoded stabilizer circuits based on the $[[5,1,3]]$ and $[[7,1,3]]$ codes \cite{Laflamme96,Steane96}, Grover search circuits, and random Clifford circuits. 
In Sec.~\ref{sec:CompareContPauli}, we introduce a model-independent matching scheme to compare noise models. A Pauli binary-symmetric channel with error rate $p$ is matched to continuous noise with spread $\sigma$ or $\kappa$. The matching uses a shared uncertainty measure: the binary entropy at readout.
The results of running these circuits with continuous Gaussian errors and comparing them with a Pauli depolarizing channel are presented in Sec.~\ref{Sec:Results-ErrorModel}.\\
We then develop an approximate analytical method for coherent error propagation, presented in Sec.~\ref{sec:approximate_method}. The method tracks small rotations through Clifford circuits without full Monte Carlo sampling. This reduces simulation cost while preserving accuracy for logical error estimates. This helps to estimate the performance of large error correcting codes under this type of non-Clifford noise without performing an exhaustive simulation.\\
In sec.~\ref{Sec:Results-AprxmModel}, we validate the approximation against brute-force simulations and identify its regime of validity. Finally, we use it to study how noise structure affects logical performance beyond raw error rates. The paper concludes with a summary and perspectives in Sec.~\ref{sec:conclusion}.

\section{Continuous Errors}
\label{sec:continuous_errors}

Continuous noise models describe gradual and random shifts in quantum gates. They differ from discrete models, which treat noise as bit flips or phase flips. Continuous errors match experimental data more closely, since real devices experience small shifts in control fields or slow drift in the environment. These shifts change the rotation angle or rotation axis of a gate and lead to coherent components that persist across circuits. Experiments in superconducting qubits show clear signs of such coherent errors~\cite{Sheldon2016,Kaufmann_2025}. Their impact differs from Pauli noise in both simulation and fault-tolerance studies~\cite{Barnes2017,Huang2019}. Models based on directional statistics, such as the von~Mises--Fisher (vMF) law and its small-angle Gaussian form, constitute a compact and physical way to describe axis and angle drift~\cite{Fisher1953,Schultz2016,Wudarski2020,Hanks2024,GeneralizedPhaseEstimation}.

Recent works have studied how such coherent unitary errors affect error correction. Gutiérrez {\it et al.}~\cite{Gutierrez2016} showed that Pauli-twirled approximations underestimate the impact of coherent errors on logical failure rates in stabilizer codes. Barnes {\it et al.}~\cite{Barnes2017} found that coherent noise leads to heavy-tailed distributions of logical failure, differing significantly from stochastic errors. Beale {\it et al.}~\cite{Beale2018} proved that repeated syndrome measurements can ``decohere'' coherent noise, effectively converting it into stochastic noise at the logical level. Cai {\it et al.}~\cite{Cai2020} demonstrated mitigation of coherent noise using Pauli conjugation techniques that outperform naive twirling. These works focus on decoding behavior, threshold shifts, or mitigation tools, whereas our approach centers on scalable modeling and entropy-matched benchmarking of noise structure across various circuit types.

\subsection{Nature of Errors}

A single-qubit state can be written on the Bloch sphere as
\begin{equation}
\ket{\psi}
=
\cos\!\left(\frac{\Theta}{2}\right)\ket{0}
+
e^{i\Phi}\sin\!\left(\frac{\Theta}{2}\right)\ket{1},
\end{equation}
where $\Theta\in[0,\pi]$ and $\Phi\in[0,2\pi]$ are the Euler angles of the quantum state on the Bloch sphere.

Single-qubit gates act as  $SU(2)$ rotations.
A general rotation takes the form
\begin{equation}
\label{generalized_error}
U(\theta,\alpha,\beta)
=
\begin{pmatrix}
e^{i\alpha}\cos\theta & i e^{i\beta}\sin\theta \\
i e^{-i\beta}\sin\theta & e^{-i\alpha}\cos\theta
\end{pmatrix},
\end{equation}
where $\theta\in[0,\pi]$ is the rotation angle and $\alpha,\beta\in[0,2\pi]$ are phase offsets.
For error modeling, we use the reduced form $U(\theta,\phi)$ with $\alpha=\beta=\phi$.
This captures small coherent miscalibrations with minimal parameters.

Noise is modeled as a deviation from the intended unitary rotation.
This deviation alters either the rotation angle, the rotation axis, or both.

\subsection{Error Model}
\label{ssec:errmodel}

Coherent noise is modeled as a random unitary rotation drawn from a probability density
on the Bloch sphere; a similar representation of errors was used by Ragazzi {\it et al.}~\cite{GeneralizedPhaseEstimation}.
The von~Mises--Fisher (vMF) distribution~\cite{Fisher1953} plays a central role.
It is the spherical analog of a multivariate normal distribution
and provides a natural description of directional uncertainty.

For a target rotation axis $\hat{\mu}$, the vMF density is
\begin{equation}
p_k(\hat{n})
=
\frac{k}{4\pi\sinh k}\, e^{k\,\hat{n}\cdot\hat{\mu}},
\qquad
\hat{n}\cdot\hat{\mu}=\cos\lambda ,
\label{eq:vMF}
\end{equation}
where $\lambda$ is the misalignment angle and $k>0$ is the concentration parameter.
Large $k$ corresponds to high gate precision.
This model has been used to characterize coherent gate errors in recent work
\cite{GeneralizedPhaseEstimation}.

In the small-angle limit $\lambda\ll1$ and for $k\gg1$,
\begin{equation}
\cos\lambda \approx 1-\frac{\lambda^2}{2},\qquad \sinh k \approx \frac{e^{k}}{2}.
\end{equation}
The distribution \eqref{eq:vMF} then reduces to a Gaussian,
\begin{equation}
p_{\sigma}(\lambda)
\approx
\frac{1}{2\pi\sigma^2}\,
e^{-\lambda^2/(2\sigma^2)},
\label{eq:gaussian_approx}
\end{equation}
with variance $\sigma^2=1/k$.

Separating polar and azimuthal components yields the bivariate Gaussian
\begin{equation}
p_{\sigma_\theta,\sigma_\phi}(\theta,\phi)
=
\frac{1}{2\pi\sigma_\theta\sigma_\phi}
\exp\!\left(
-\frac{\theta^2}{2\sigma_\theta^2}
-\frac{\phi^2}{2\sigma_\phi^2}
\right),
\label{eq:2d_gaussian}
\end{equation}
valid for small and near-isotropic noise.

Experimental data from superconducting qubits shows stable rotation biases
with clear directional structure
\cite{Kaufmann_2025}.
These observations agree with the vMF model and its Gaussian limit
in Eqs.~\eqref{eq:gaussian_approx} and~\eqref{eq:2d_gaussian}.

To apply this model at the circuit level, the circuit is decomposed into Clifford gates.
Three assumptions are made:
\begin{enumerate}
    \item Only single-qubit gates induce coherent noise.
    \item CNOT gates are noiseless but propagate existing errors.
    \item The noise is isotropic, with $\sigma_\theta=\sigma_\phi=\sigma$.
\end{enumerate}

Algorithm~\ref{alg:coherent_noise_simulation} summarizes the simulation procedure.
After each single-qubit gate, two independent angles are sampled
to construct a small coherent rotation using Eq.~\eqref{generalized_error}.
These rotations are inserted into the circuit.
The noisy circuit is then executed on a chosen simulator.
Repeating this process yields an ensemble of outcomes
from which probability distributions or state vectors are obtained.

\begin{algorithm}[t!]
   \begin{algorithmic}
   \State Choose circuit $C$, noise grid $\Sigma=\{\sigma\}$,
   number of circuit instances $N_c$, and \textit{Simulator}
   \For{$\sigma \in \Sigma$}
        \State Initialize \textbf{results} $\gets \varnothing$
        \For{$i = 1, \ldots, N_c$}
            \State Set $C' \gets C$
            \ForAll{single-qubit gate $G$ in $C'$}
                \State Sample $(\theta,\phi)\sim\mathcal{N}(0,\sigma^2)$ \Comment{Eq.~\eqref{eq:2d_gaussian}}
                \State Insert $U_{\mathrm{err}}(\theta,\phi,\phi)$ after $G$
                \Comment{Eq.~\eqref{generalized_error}}
            \EndFor
            \State Execute $C'$ on \textit{Simulator}
            \State Append outcomes to \textbf{results}
        \EndFor
   \EndFor
   \State Return logical error curve
   $\epsilon_{\text{logical}}(\sigma)$
   \end{algorithmic}
   \caption{Circuit simulation with coherent Gaussian noise.}
   \label{alg:coherent_noise_simulation}
\end{algorithm}

\section{Quantum Circuits Studied}
\label{QuantumCircuitsStudied}

\subsection{Stabilizer Code Circuits}
\label{ssec:CodeCircs}

\begin{figure}[t]
    \centering
    \includegraphics[width=0.65\linewidth]{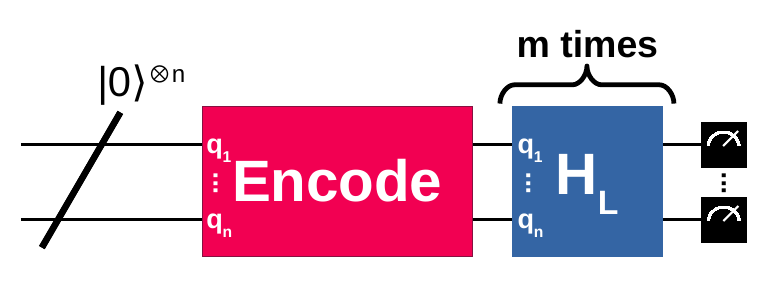}
    \caption{ \change{Encoding operation on $n$ qubits, followed by $m$ consecutive of $H_L$ circuits to increase the depth of the circuit.}}
     \label{generalcircuitec=0}
\end{figure}

\begin{figure}[t]
    \centering
    \includegraphics[width=\linewidth]{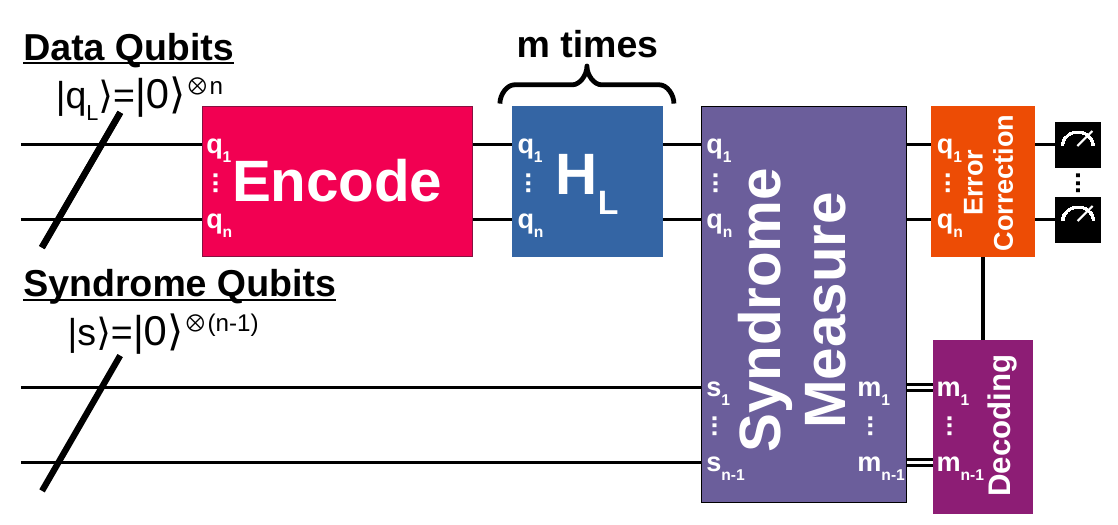}
    \caption{Encode one logical qubit of any code, apply a series of logical Hadamards and a round of syndrome extraction, and measure the decoded state. Double lines represent classical bits that hold qubit information after measurement to be processed in the decoder.}
    \label{generalcircuitec=1}
\end{figure}

First, we study circuits based on the $[[5,1,3]]$ and $[[7,1,3]]$  stabilizer codes from Refs.~\cite{Laflamme96,Steane96}. In general, a stabilizer code of the form $[[n,k,d]]$ encodes $k$ logical qubits into $n$ physical qubits and can correct up to $\lfloor (d-1)/2 \rfloor$ arbitrary single-qubit errors on each encoded logical qubit. These are minimal codes that correct any single-qubit error, and are standard examples within the stabilizer formalism \cite{Gottesman1997,Terhal2015,BRADSHAW2026170353}. \change{The circuit structure used throughout this work is shown in Figs.~\ref{generalcircuitec=0} and ~\ref{generalcircuitec=1}. These circuits begin by encoding $n$ physical qubits into an entangled state, the logical qubit in the case of Fig.~\ref{generalcircuitec=1}, followed by the sequential application of multiple logical Hadamard gates or circuits to amplify the effect of noise. Fig.~\ref{generalcircuitec=1} concludes with a single round of syndrome measurement, decoding, and error correction}, as described in standard quantum error-correction protocols \cite{Terhal2015,BRADSHAW2026170353}. The final state of the data qubits is then measured and classically post-processed.

\begin{figure}[t]
    \centering
    \includegraphics[width=\linewidth]{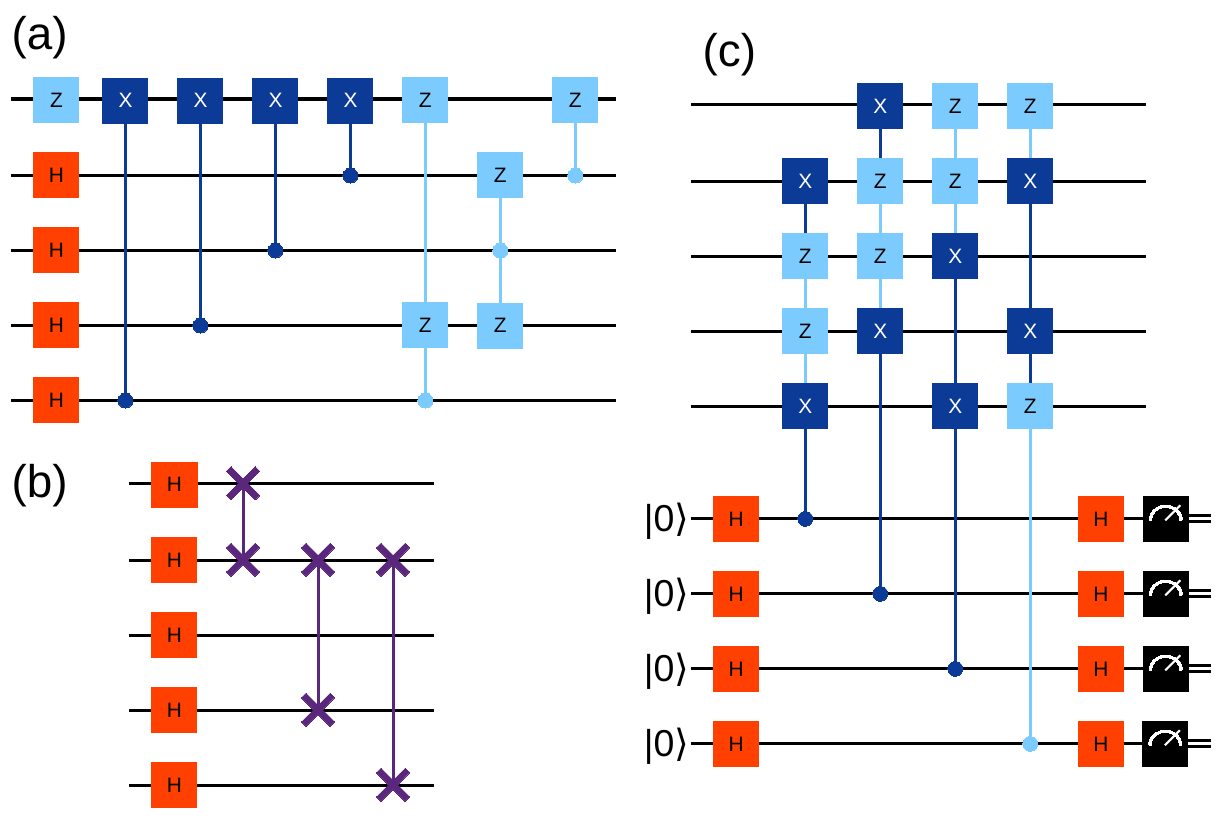}
    \caption{Circuits for the $[[5,1,3]]$ code: (a) encoding circuit, (b) logical Hadamard circuit \cite{Yoder2016}, (c) syndrome-measurement circuit.}
    \label{fig:513_circuits}
\end{figure}

\begin{figure}[t]
    \centering
    \includegraphics[width=\linewidth]{Figures/713circuits.pdf}
    \caption{Circuits for the $[[7,1,3]]$ code: (a) encoding circuit \cite{Zhao2020}, (b) logical Hadamard circuit, (c) syndrome-measurement circuit.}
    \label{fig:713_circuits}
\end{figure}

Figs.~\ref{generalcircuitec=0} and ~\ref{generalcircuitec=1} encode one logical qubit in a non-fault-tolerant manner, as shown in Figs.~\ref{fig:513_circuits}a and~\ref{fig:713_circuits}a. 
The notations for gates and circuits are from the standard literature, such as Ref.~\cite{Nielsen_Chuang_2010}.
This encoding method applies single- and two-qubit gates between the qubits within the same block (the physical qubits of the encoded qubit), resulting in a non-transversal circuit, which in turn is non-fault-tolerant \cite{Terhal2015}. After encoding, a sequence of $m$ logical operations is applied within the code space. To probe noise accumulation with circuit depth, $m$ consecutive logical Hadamard gates are applied.

The logical Hadamard exchanges the logical computational and $X$ bases,
\begin{equation}
H_L\ket{0_L;1_L}=\ket{+_L;-_L},
\qquad
H_L\ket{+_L;-_L}=\ket{0_L;1_L},
\end{equation}
where $\ket{\pm_L} = (\ket{0_L} \pm \ket{1_L})/\sqrt{2}$. Implementations of the logical Hadamard for both codes are shown in Figs.~\ref{fig:513_circuits}b and~\ref{fig:713_circuits}b, respectively. For the $[[5,1,3]]$ code, the logical Hadamard requires a combination of physical Hadamard gates and SWAP operations, making it non-transversal. For the $[[7,1,3]]$ code, the logical Hadamard consists solely of physical Hadamard gates, and is therefore transversal.

After the logical operations, the circuit in Fig.~\ref{generalcircuitec=1} applies one round of error correction. A single syndrome-extraction and correction cycle is performed. For the circuit in Fig.~\ref{generalcircuitec=1}, the information of data qubits is directly extracted, errors are left to accumulate as more logical gates are applied without detection or correction.

Syndromes are obtained using circuit-based stabilizer measurements. Each stabilizer couples the data qubits to a single ancilla through CNOT and CZ gates, as shown in Figs.~\ref{fig:513_circuits}c and~\ref{fig:713_circuits}c. Measured syndromes are decoded using a hard-decision lookup table that maps each outcome to the corresponding physical error to be then reversed by error correction.

Finally, the encoded block is measured. Outcomes being one of the superposed states of $\ket{0_L}$ or $\ket{1_L}$ are recorded as logical $\ket{0}$ or $\ket{1}$. This setup provides a controlled framework to study how continuous and Pauli noise affect the correctness of the encoded information under realistic gate errors.

\change{
Throughout this section, we use the probability of not measuring the expected output state as the main performance metric. For the uncoded stabilizer circuit, shown in Fig.~\ref{generalcircuitec=0}, we denote this quantity by probability of error \(p_{\mathrm{err}}\). It measures the output error produced by the noisy circuit.

For the circuit with syndrome extraction and recovery, shown in Fig.~\ref{generalcircuitec=1}, we denote the same type of output failure probability by logical error rate \(r_{\mathrm{logical}}\). In this case, the remaining error is interpreted as a logical error after correction. Thus,
\begin{equation}
\label{QEC:metric}
p_{\mathrm{err}} = 1 - P_{\mathrm{expect}},
\qquad
r_{\mathrm{logical}} = 1 - P_{\mathrm{expect}},
\end{equation}
where \(P_{\mathrm{expect}}\) is the probability of obtaining the expected output state.

The two quantities are computed in the same way, but they refer to different circuits. The quantity \(p_{\mathrm{err}}\) describes the error accumulated at the output of the uncoded circuit. The quantity \(r_{\mathrm{logical}}\) describes the performance of the error correcting code under some noise model.
}

\subsection{Grover Search Circuits}
\label{subsec:GSACircuit}

\begin{figure}[t!]
    \centering
    \includegraphics[width=\linewidth]{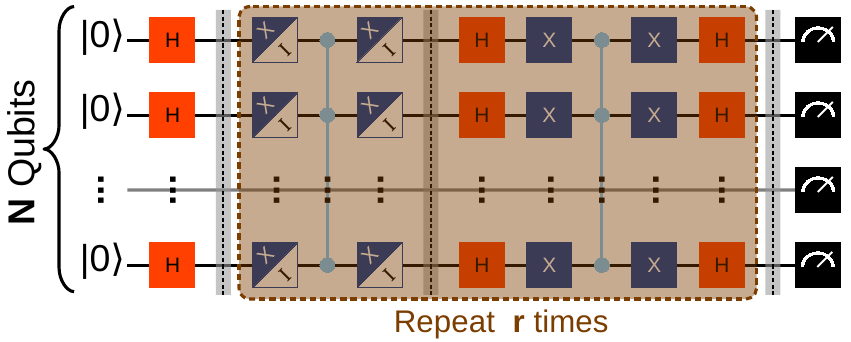}
    \caption{A general Grover's search circuit \cite{elkaderi2023}.}
    \label{generalcircuitGSA}
\end{figure}

Grover’s search algorithm \cite{Grover96,Grover97} is also studied under both noise models. This algorithm acts on $N$ qubits and augments the probability of measuring one of the states $\ket{\omega}$ of a Hilbert space of dimension $2^N$. A generalized circuit structure is shown in Fig.~\ref{generalcircuitGSA}.

The circuit begins by preparing all qubits in the $\ket{+}$ state. The shaded block in Fig.~\ref{generalcircuitGSA} consists of two components. The first is a phase oracle that marks the target state. It is implemented using a multi-control $Z$ gate $C^{N-1}Z$~\cite{Nielsen_Chuang_2010} surrounded by $X$ gates chosen according to the marked state.

The second component is the diffuser. It applies a phase shift to all computational basis states. Its structure mirrors that of the oracle, with a $C^{N-1}Z$ gate placed between layers of Hadamard and $X$ gates. The shaded block is repeated
\begin{equation}
\label{Eq:GSArepetitions}
r = \left\lfloor \frac{\pi N}{2} - \frac{1}{2} \right\rfloor
\end{equation}
times to obtain a closer estimate of the required value. Grover’s search algorithm is probabilistic, with a success probability that depends on the number of Grover iterations. When the algorithm is executed with the optimal (integer) number of iterations, the residual algorithmic error due to the discretization of the rotation angle scales as
\begin{equation}
    \label{Eq:gsafail}
    P_{\rm fail} = O\!\left(2^{-N}\right) ,
\end{equation}
where $N$ is the number of qubits encoding the search space.
These circuits provide a structured test of noise accumulation in multi-qubit algorithms.

\change{The $C^{N-1}Z$ gate in Fig.~\ref{generalcircuitGSA} is shown as a compact multi-controlled gate. In the simulations below, it is decomposed into single- and two-qubit gates before the noise model is applied.}

Recent studies also assess Grover circuits under various noise conditions.
Kumar~\textit{et al.}~\cite{Kumar2025GroverNoise} compare the degradation in Grover's success probability under different types of noise, including amplitude damping and depolarizing noise, and propose circuit designs with higher resilience. 
Ijaz~\textit{et al.}~\cite{Ijaz2023} analyze Grover’s algorithm using both simulated and theoretical error models and show that coherent noise sources can delay convergence or lower peak fidelity, depending on the iteration count.

Again, in this section, performance is quantified using the metric
\begin{equation}
\label{GSA:metric}
p^{\mathrm{GSA}}_{\mathrm{err}} = 1 - p_{\ket{\omega}} ,
\end{equation}
where $p_{\ket{\omega}}$ is the probability of the target state $\ket{\omega}$. This represents the probability of not measuring the targeted state.

\subsection{Random Clifford Circuits}
\label{Sec:RandomCircs}

Random Clifford circuits are used as a complementary benchmark for the approximation model in Sec.~\ref{sec:approximate_method}.  
Each circuit acts on $n$ qubits and is built from Hadamard and CNOT gates. Gate locations are sampled uniformly at random, while the total gate count is fixed. Each instance contains a set number of Hadamards and CNOTs.

Clifford circuits form a class of dynamics that is efficiently simulable while still generating nontrivial entanglement and operator spreading \cite{Bravyi2019simulationofquantum}.  
More general random circuit models are often used to study universal features of quantum dynamics and entanglement growth \cite{Fisher2023}.
A recent investigation analyzes the statistical physics properties of noisy random circuits \cite{Sauliere2026}.

Here we pursue a different goal.
These circuits are not constructed to encode information or protect against noise. They are not treated as a physical model of interest. They are used only as controlled test cases.

Their role is to probe how errors propagate under different gate orderings, depths, and qubit couplings. Randomization removes structure that could hide systematic effects. What remains is a direct test of how the approximation handles coherent error buildup.

No ensemble properties are analyzed. We do not extract scaling laws or typical behavior. Each circuit is treated as a separate instance. The circuits act as stress tests. They reveal how errors spread and mix across qubits through repeated Clifford layers.

Random Clifford circuits appear widely in quantum error correction. They are used to generate stabilizer codes and study decoding thresholds. Low-depth constructions perform well under Pauli noise
\cite{BrownFawzi2015,Gullans2021,Darmawan2024}.

Miller~\textit{et al.}~\cite{miller2025efficientsimulationcliffordcircuits} gave an efficient method to simulate Clifford circuits with Markovian noise and showed that structured noise maps can mimic correlated coherent channels. Jin~\textit{et al.}~\cite{JinSung2021} used hardware-efficient Clifford circuits to distinguish noise types in multiqubit devices.

Here the use is narrower. The circuits are not studied for their own properties. They are used only to compare full coherent-noise simulation with the approximation model. Agreement across many random instances supports the validity of the approximation.

\section{Entropy-Matched Comparison of Noise Models}
\label{sec:CompareContPauli}

In this section, we evaluate coded and uncoded circuits, as well as Grover search circuits from Sec.~\ref{QuantumCircuitsStudied} under two noise models: a discrete Pauli channel and continuous small-angle coherent rotations.
To enable a fair comparison, both models are mapped to an effective binary-symmetric channel (BSC) at readout, and the comparison is carried out at matched binary entropy.
This approach removes scale dependence and isolates the role of noise structure.

Matched-entropy comparisons have been widely used in the literature to distinguish the impact of coherent versus stochastic noise.
Gutiérrez and Brown showed that coherent unitary errors can cause significantly higher logical error rates than their entropy-equivalent Pauli-twirled counterparts in QEC circuits, especially below threshold~\cite{Gutierrez2016}.
Barnes {\it et al.}\ reported that coherent errors generate heavy-tailed logical failure distributions in stabilizer codes, unlike narrow distributions seen under Pauli noise~\cite{Barnes2017}.
Beale {\it et al.}\ mathematically demonstrated that syndrome measurements progressively decohere physical coherent errors into effective stochastic errors at the logical level, providing partial justification for entropy-based comparisons~\cite{Beale2018}.

Bravyi {\it et al.}~\cite{Bravyi2018} extended this analysis to surface codes and found that coherent errors interact with the code structure in nontrivial ways. Their work showed that Pauli twirling can underestimate logical error rates in low-noise regimes, reinforcing the importance of matched comparisons between noise types.

Entropy matching has also been applied to Clifford and algorithmic circuits.
Studies show that coherent errors can accumulate constructively in Clifford layers, leading to quadratic error growth in depth, while Pauli noise accumulates linearly~\cite{Huang2019}.
In algorithm benchmarks, Ijaz and Faryad found that Grover’s search is more sensitive to depolarizing noise than to small coherent over-rotations, again under entropy-matched settings~\cite{Ijaz2023}.
Similarly, Kumar {\it et al.}~\cite{Kumar2025GroverNoise} demonstrated that Grover performance degrades differently under distinct noise models and highlighted circuit constructions that maintain robustness under coherent noise.

Together, these findings confirm that binary entropy is a practical and meaningful calibration tool when comparing the structure-specific impacts of different noise types.

\subsubsection{Pauli Noise (Reference Model)}

We consider a symmetric Pauli channel with total error rate $p$.
Each of $X$, $Y$, and $Z$ occurs with probability $p/3$.
The gate acts ideally with probability $1-p$.
A $Z$-basis measurement yields a BSC with bit-flip (bf) probability
\begin{equation}
p_{\mathrm{bf}}^{\mathrm{Pauli}}=\frac{2p}{3},
\end{equation}
since $X$ and $Y$ flip the outcome.
The corresponding binary entropy~\cite{Richardson_Urbanke_2008} is
\begin{align}
H_{\mathrm{Pauli}}(p)
&= h_2\!\left(\frac{2p}{3}\right), \nonumber\\
h_2(x)
&= -x\log_2 x-(1-x)\log_2(1-x).
\label{eq:H-pauli}
\end{align}

\subsubsection{Continuous rotations: von~Mises--Fisher (vMF)}

Coherent control errors are modeled as random tilts of the rotation axis.
Let $\lambda$ denote the angle between the intended and actual axes.
\change{We use this tilt to define an effective binary error at readout.}
A $Z$-basis bit flip occurs with probability
$\sin^2(\lambda/2)=\tfrac12(1-\cos\lambda)$.
The effective flip rate follows from averaging this quantity over the noise distribution.

For axis fluctuations drawn from the von~Mises--Fisher (vMF) distribution,
the mean value of $\cos\lambda$
\change{is the average projection of the sampled axis onto the intended axis. It}
is given by the Langevin function
\begin{equation}
\mathcal L(\kappa)=\coth\kappa-\frac{1}{\kappa},
\end{equation}
where $\kappa$ is the concentration parameter. The resulting crossover probability is
\begin{equation}
p_{\mathrm{bf}}^{\mathrm{vMF}}(\kappa)
=\frac12\!\left(1-\mathcal L(\kappa)\right).
\end{equation}
We map the continuous error spectrum to an effective binary variable
\change{at readout, namely the probability of obtaining the wrong outcome in the measurement basis.}
This enables comparison with the binary symmetric Pauli channel. Substituting the resulting crossover probability into Eq.~\eqref{eq:H-pauli} yields the binary entropy
\begin{equation}
H_{\mathrm{vMF}}(\kappa)
= h_2\!\bigl(p_{\mathrm{bf}}^{\mathrm{vMF}}(\kappa)\bigr).
\end{equation}
Increasing $\kappa$ improves axis alignment and suppresses errors. 

In the narrow-angle limit $\kappa\gg1$, the vMF distribution reduces to an isotropic
Gaussian in the tangent plane.
Using $\mathcal L(\kappa)\approx1-\kappa^{-1}$ yields
\begin{equation}
p_{\mathrm{bf}}^{\mathrm{vMF}}(\kappa)\approx\frac{1}{2\kappa}.
\end{equation}
If $\sigma^2$ denotes the variance of each tangent-plane Gaussian component, then
$\langle\lambda^2\rangle\simeq 2\sigma^2$. Since the narrow-angle vMF limit gives
$\langle\lambda^2\rangle\simeq 2/\kappa$, we identify $\sigma^2\simeq 1/\kappa$.
This gives
\begin{equation}
    p_{\mathrm{bf}}\simeq \frac{\langle\lambda^2\rangle}{4} \simeq \frac{\sigma^2}{2}.
\end{equation}

We also tested a hard-threshold model where errors are counted only beyond a fixed angular cutoff. This leads to a lower binary entropy at the same $\kappa$, hence a less conservative estimate of noise. For this reason, we do not use it in the main analysis.

\section{Results and Discussion: Continuous versus Pauli Noise}
\label{Sec:Results-ErrorModel}

This section compares circuit performance under continuous coherent noise and Pauli noise.
The study covers stabilizer-code circuits and Grover search circuits.
All simulations follow Algorithm~\ref{alg:coherent_noise_simulation} and use Monte Carlo sampling of the noise.
Each run produces a probability distribution over measurement outcomes.

\subsection{$[[5,1,3]]$ and $[[7,1,3]]$ Circuit}
\label{Sec:CodeResConPauli}

\begin{figure}[t!]
    \centering
    \includegraphics[width=\linewidth]{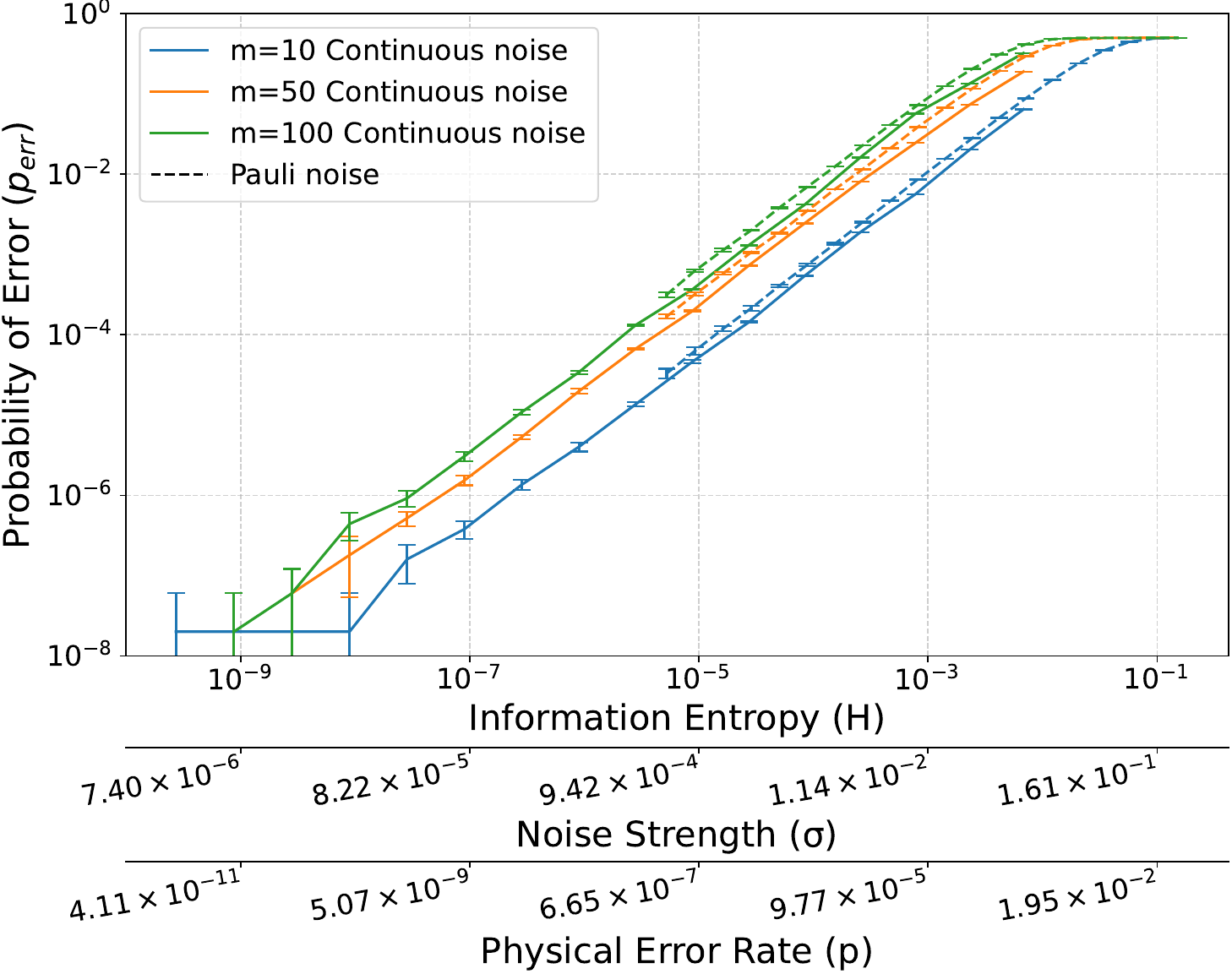}
    \caption{Probability of error $p_{\mathrm{err}}$ for the circuit in Fig.~\ref{generalcircuitec=0} with the $[[5,1,3]]$ circuits of Fig.~\ref{fig:513_circuits}. Solid curves correspond to continuous noise. Dashed curves correspond to Pauli noise. The circuit of Fig.~\ref{generalcircuitec=0} is run with $m=10,50,100$ logical Hadamard circuits. The bottom, middle, and top axes show the Pauli error rate $p$, continuous noise strength $\sigma$, and matched binary entropy $H$.
}
    \label{fig:resullt11}
\end{figure}

\begin{figure}[t!]
    \centering
    \includegraphics[width=\linewidth]{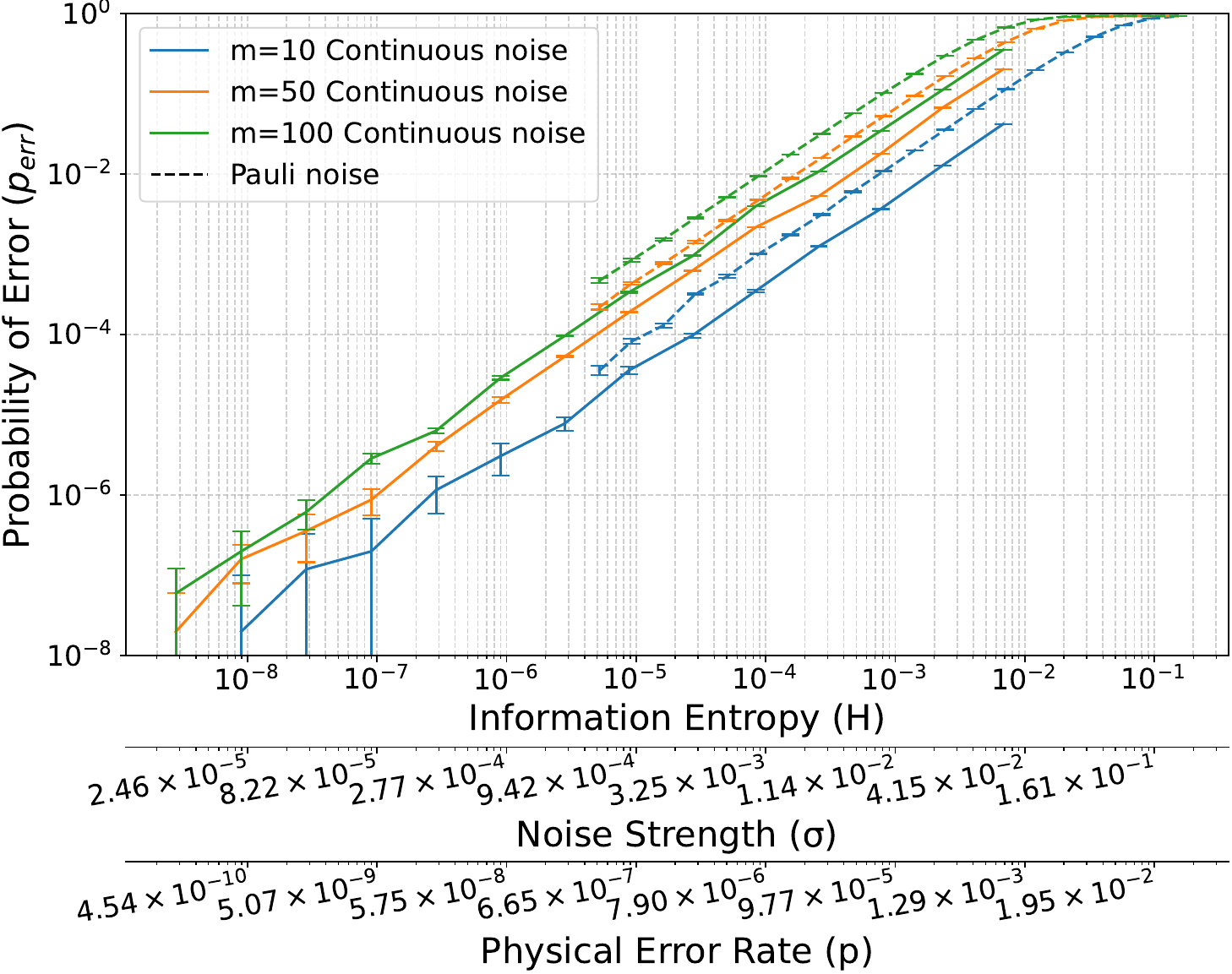}
    \caption{Same as Fig.~\ref{fig:resullt11} but for the $[[7,1,3]]$ circuits of Fig.~\ref{fig:713_circuits}.
}
    \label{fig:resullt12}
\end{figure}

\begin{figure}[t!]
    \centering
    \includegraphics[width=\linewidth]{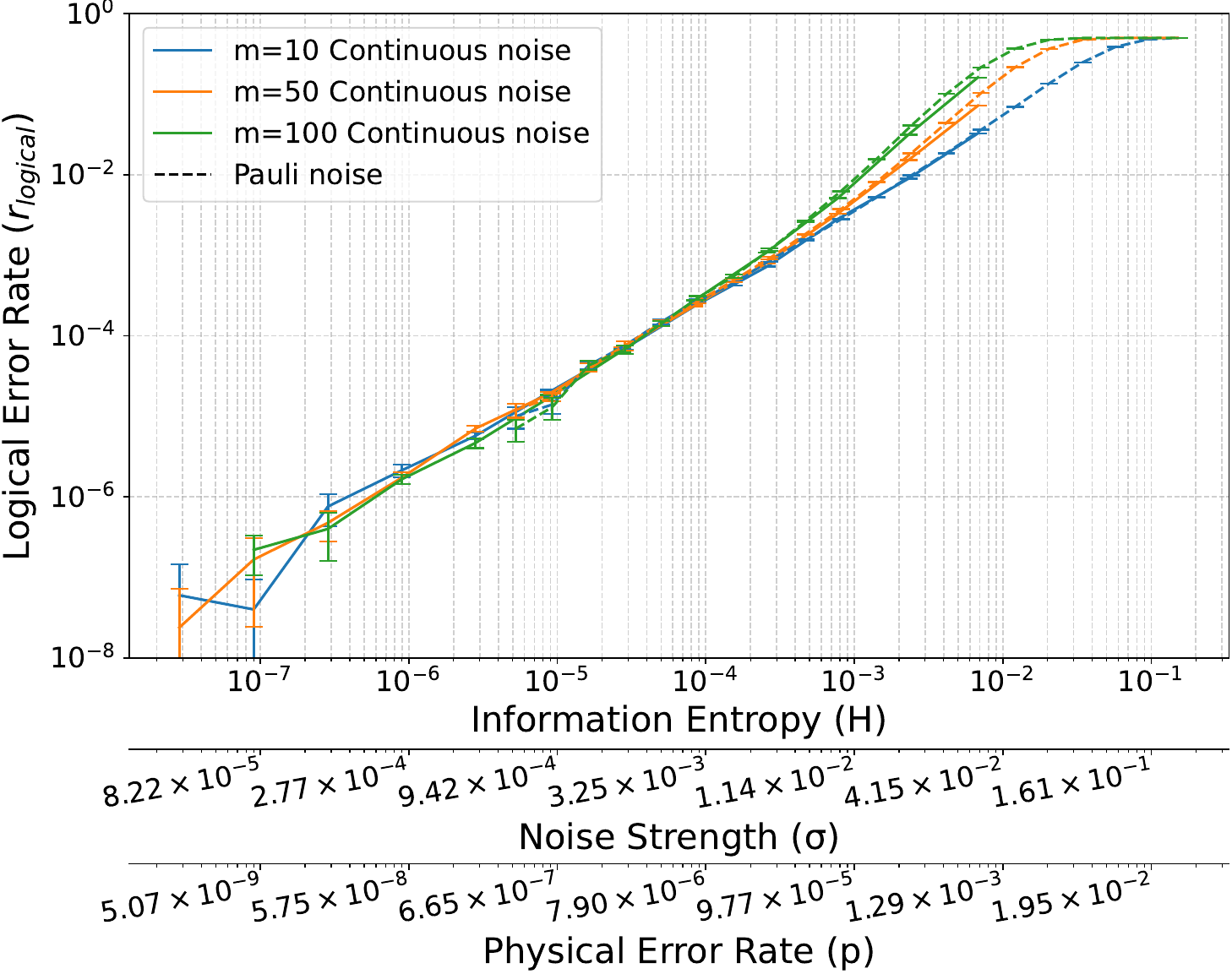}
    \caption{Logical error rate $r_{\rm logical}$ for the circuit in Fig.~\ref{generalcircuitec=1} with the $[[5,1,3]]$ circuits of Fig.~\ref{fig:513_circuits}.
}
    \label{fig:resullt21}
\end{figure}

\begin{figure}[t!]
    \centering
    \includegraphics[width=\linewidth]{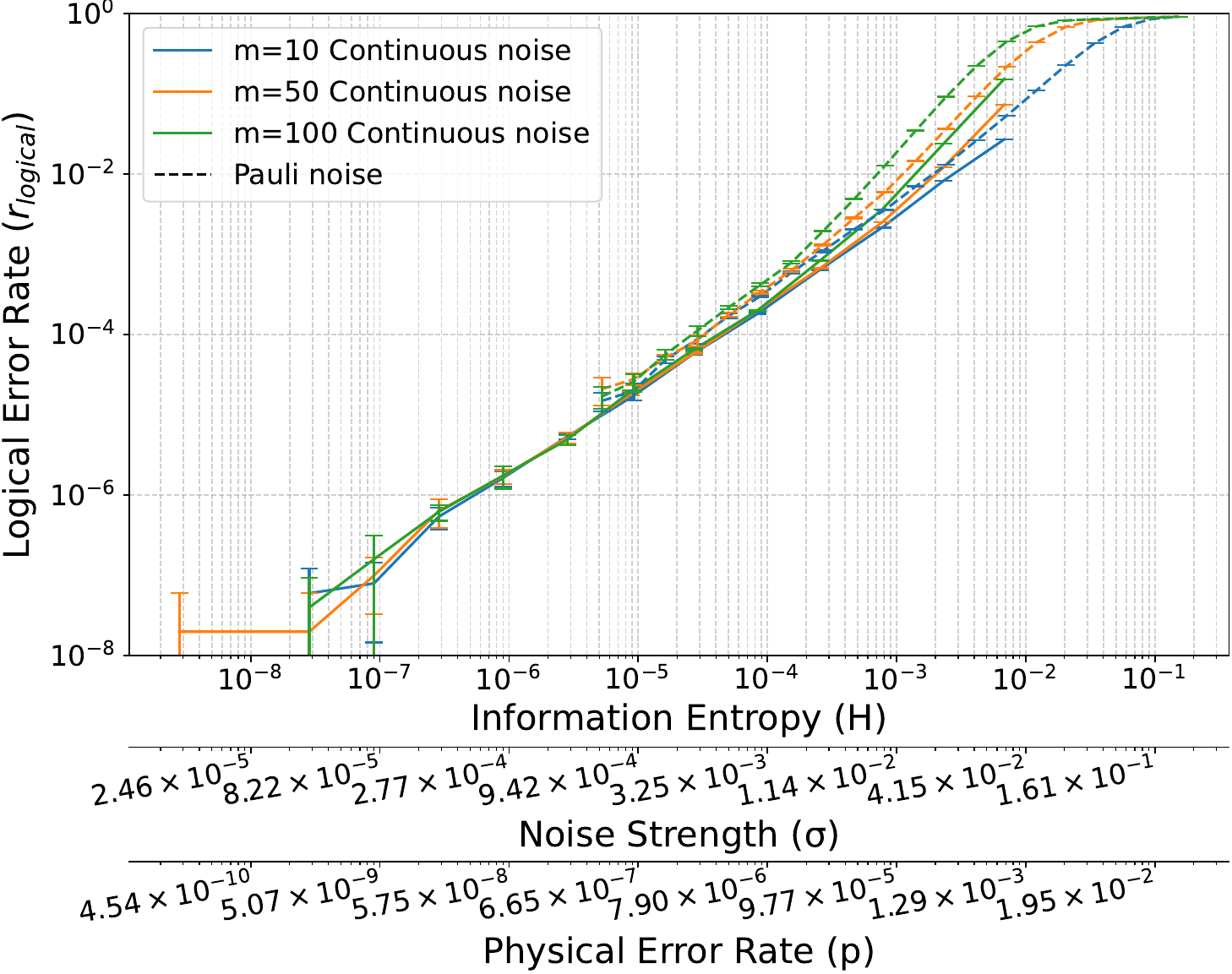}
    \caption{Same as Fig.~\ref{fig:resullt21} but for the $[[7,1,3]]$ circuits of Fig.~\ref{fig:713_circuits}.
}
    \label{fig:resullt22}
\end{figure}

For each parameter setting, results from all realizations are aggregated into one distribution.
The error probability $p_{\mathrm{err}}$ and the logical error rate $r_{\rm logical}$ are computed using Eq.~\eqref{QEC:metric} for the circuits in Figs.~\ref{generalcircuitec=0} and ~\ref{generalcircuitec=0} respectively, where $p_{\mathrm{err}}$ is shown on the vertical axes of Figs.~\ref{fig:resullt11} and~\ref{fig:resullt12}, and $r_{\rm logical}$ is shown on the vertical axes of Figs.~\ref{fig:resullt21} and~\ref{fig:resullt22}.
Each data point uses $O(10^3)$ independent circuit instances.
Every instance is executed with $100$ measurement shots.

The circuits in Figs.~\ref{generalcircuitec=1} and~\ref{generalcircuitec=0} are evaluated using the circuits of both codes presented in Figs.~\ref{fig:513_circuits} and ~\ref{fig:713_circuits}. 
The number of logical Hadamard gates is set to $m=10,50,100$.
The results appear in Figs.~\ref{fig:resullt11},~\ref{fig:resullt12},~\ref{fig:resullt21} and~\ref{fig:resullt22}.

Two noise models are considered.
The first is continuous coherent noise with strength $\sigma$.
The second is a symmetric Pauli channel with physical error rate $p$.
Each figure includes three horizontal axes.
The bottom axis shows $p$,
the middle axis shows the matching range of $\sigma$, and
the top axis reports the corresponding binary entropy.
Note that actual data for Pauli errors is in the range $p \in[10^{-6}/3,10^{-1}/3]$.

The entropy axis, defined in Sec.~\ref{sec:CompareContPauli}, enables direct comparison.
Both noise models are matched at equal readout uncertainty.
This isolates the effect of noise structure on logical performance.

Figures~\ref{fig:resullt11} and~\ref{fig:resullt12} show the effect of circuit depth and error accumulation. When no error correction
is applied at the end of the circuit, the logical error probability increases with the number $m$ of repeated logical Hadamard circuits, for both noise models. This is expected, since deeper circuits contain noisier gates and therefore accumulate more errors.

When
error correction is applied, the logical error probability is reduced, see Figs.~\ref{fig:resullt21} and~\ref{fig:resullt22}. At low physical error strength, the curves approach a common asymptotic behavior for both noise models. This shows that the error-correction scheme suppresses errors introduced by noisy logical Hadamard gates. For larger values of $m$, this low-error regime is reached only at lower physical error strengths (a longer tail). In other words, deeper circuits require weaker physical noise to reach the same logical performance as shallower circuits.

Another
point that can be observed in Figs.~\ref{fig:resullt11},~\ref{fig:resullt12},~\ref{fig:resullt21} and~\ref{fig:resullt22} is that the continuous-error curves and the Pauli-error curves show similar trends.
This remains true when
error correction is applied as shown in Figs.~\ref{fig:resullt21} and~\ref{fig:resullt22}. Thus, the syndrome-extraction circuit and the decoder used in Sec.~\ref{ssec:CodeCircs} treat the two noise models similarly at the logical level.
In fact, the curves for continuous and Pauli noise in the $[[5,1,3]]$ code are very close in
Fig.~\ref{fig:resullt21} after error
correction and with matched binary entropies, whereas there is a clearly visible difference for the $[[7,1,3]]$ code, compare
Fig.~\ref{fig:resullt22}.

This similar behavior of the noise models can be understood from the Pauli expansion of a single-qubit coherent error. Any single-qubit rotation error can be written as
\begin{equation}
\label{eq:RotinPauli}
    R =
    c_I I + c_X X + c_Y Y + c_Z Z,
    \qquad c_I,c_X,c_Y,c_Z\in \mathds{C}.
\end{equation}
For the error model in Eq.~\eqref{generalized_error}, the coefficients are
\begin{align}
    c_I &= \cos\theta \cos\phi, \\
    c_X &= i\sin\theta \cos\phi, \\
    c_Y &= -i\sin\theta \sin\phi, \\
    c_Z &= i\cos\theta \sin\phi .
\end{align}
After the syndrome circuit, the ancilla state contains the syndrome information associated with the Pauli components in Eq.~\eqref{eq:RotinPauli}. Measurement of the syndrome projects the state onto one of these Pauli-error subspaces. The data qubits are then, in turn, projected to a state with a definite syndrome, which the decoder maps to a correction. This explains why, in these error-correction circuits and for matched binary entropies, continuous coherent errors exhibit an effective behavior close to that of Pauli errors after syndrome extraction.

\subsection{Grover Search Circuit}

\label{ssec:gsaresults}

Next, we evaluate
Grover search circuits using the construction in Fig.~\ref{generalcircuitGSA}.
The algorithm is described in Sec.~\ref{subsec:GSACircuit}.
Both Pauli and continuous noise models are applied.

\begin{figure}[t!]
    \centering
    \includegraphics[width=\linewidth]{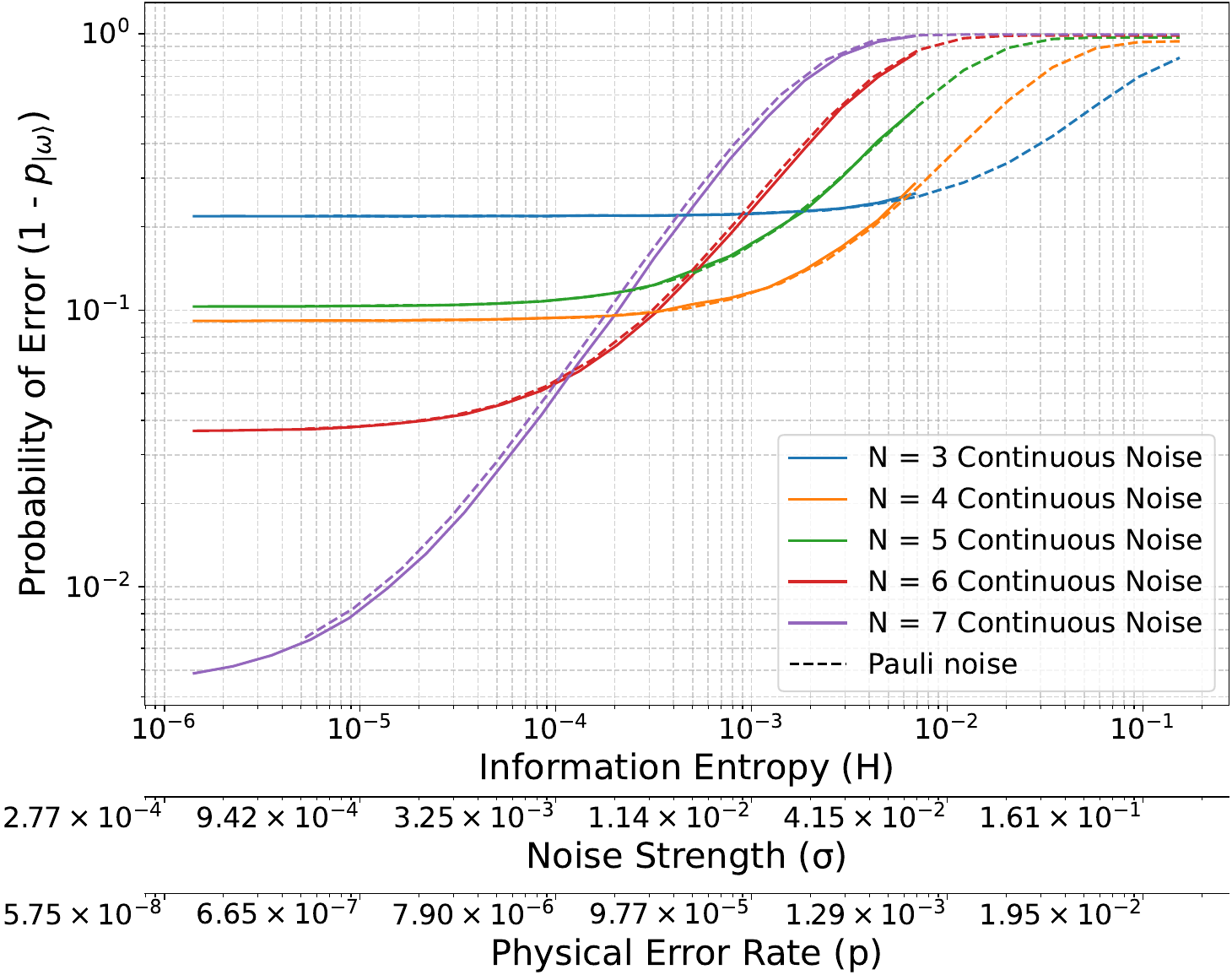}
    \caption{Error probability of Grover search circuits for $N=3,4,5,6,7$ qubits.
    Solid curves show continuous noise.
    Dashed curves show Pauli noise.}
    \label{fig:resulltGSA}
\end{figure}

The circuit is simulated for $N=3,4,5,6,7$ qubits.
We plot the error probability defined in Eq.~\eqref{GSA:metric} against noise strength.
Results are shown in Fig.~\ref{fig:resulltGSA}.

At low noise levels, circuits with more qubits perform better, as mentioned in Sec.~\ref{Sec:RandomCircs},
see Eq.~\eqref{Eq:gsafail}.
The success probability increases with $N$.
This follows directly from Grover’s amplification mechanism.
Applying the optimal number of iterations in Eq.~\eqref{Eq:GSArepetitions} increases overlap with the target state $\ket{\omega}$.

As noise increases, a crossover appears.
Smaller circuits outperform larger ones.
Noise overwhelms the algorithmic gain.
Scaling without correction or mitigation becomes ineffective.
This behavior agrees with earlier results~\cite{elkaderi2023}.

At matched entropy, Grover search behaves similarly under the two noise models. The Pauli and continuous-noise curves follow the same overall trend in Fig.~\ref{fig:resulltGSA}, with no clear systematic separation between them. This is notable because the Grover circuit is not a Clifford circuit; its oracle and diffusion steps include multi-controlled phase operations, so its noise response is not expected to follow directly from stabilizer-code behavior. Still, the result is consistent with the corrected stabilizer-code results of Sec.~\ref{Sec:CodeResConPauli}, where the Pauli and continuous curves are also found to be comparable.

One possible explanation is the final measurement basis. In these simulations, the output state is measured in the computational ($Z$) basis. This readout is mainly sensitive to errors that change the measured bit value. Thus, the part of a continuous rotation that acts like an ($X$)-type error can appear as a bit flip, while phase-like components can have a weaker direct effect on the measured probability. At the level of the output metric, this can make the continuous channel look closer to a Pauli channel than it is at the gate level.

\section{Approximate Treatment of the Noise}

\label{sec:approximate_method}

Accurately simulating the full coherent noise model described in Section~\ref{ssec:errmodel} is computationally expensive because it requires generating and executing a large ensemble of randomly perturbed circuits (Algorithm~\ref{alg:coherent_noise_simulation}). To reduce this complexity and enable scalable analysis of larger quantum circuits, we introduce an approximate propagation method that propagates \emph{error distributions} through the circuit, rather than sampling individual error instances.

Recent studies suggest that coarse descriptions of noisy circuits can capture dominant noise effects. In highly random circuits, local noise tends to spread across many qubits and can approach an effective global noise model \cite{Dalzell2024,Deshpande2022,Foldager_2024}. Other works analyze how circuit depth and randomness influence circuit complexity and state structure \cite{Haferkamp2022,Suzuki2025quantumcomplexity}. It has also been shown that suitable gate ensembles or twirling operations can transform structured noise into effective depolarizing channels \cite{Tsubouchi2025}. At the same time, more realistic noise models may retain structure in the output distribution and deviate from simple white-noise descriptions \cite{Fefferman2024}. These results motivate simplified approaches that track the dominant effect of noise without resolving every error instance.

The core idea is to describe each qubit by a probability distribution $p(\vec r)$ over the Bloch sphere—capturing uncertainty from accumulated coherent errors—rather than by a single Bloch vector $\vec r$. Under the assumption of small, isotropic Gaussian errors ($\sigma\ll 1$), these distributions propagate \emph{analytically} through the circuit with simple update rules.

This approach addresses the high cost of coherent-noise simulations by leveraging the analytical structure of small-angle error accumulation. While efficient simulation strategies have been developed for stochastic models—such as the stabilizer-based techniques in Ref.~\cite{miller2025efficientsimulationcliffordcircuits}—our method handles directional uncertainty in coherent noise through deterministic propagation of error distributions. By avoiding full Monte Carlo sampling, it enables scalable and structure-aware analysis of noise effects in Clifford circuits under coherent perturbations.

\subsection{Propagation Through Single-Qubit Gates}
\label{ssec:propagation_single_qubit}
\change{
Each qubit carries a local Gaussian error described by the angular widths
$(\sigma_\theta,\sigma_\phi)$ of the distribution
$p_{\sigma_\theta,\sigma_\phi}(\theta,\phi)$
[Eq.~\eqref{eq:2d_gaussian}].
In the approximation used here, the errors are not inserted after every gate.
Instead, each qubit stores the variances of one final effective error.
These variances are initialized to zero and are updated as the circuit is read gate by gate.

A faulty single-qubit gate applies a fresh Gaussian noise on top of the
Gaussian error already carried by the qubit. Since the two errors are
independent, the new distribution is their convolution. In the small-angle
regime, this keeps the distribution Gaussian and only changes its variance.
We take the noise to be symmetric in the two angular directions, with the same
strength $\sigma$ for both $\theta$ and $\phi$. Thus,
\begin{equation}
v_{\theta,\mathrm{new}}=v_\theta+\sigma^2,
\qquad
v_{\phi,\mathrm{new}}=v_\phi+\sigma^2 ,
\label{eq:single_qubit_variance_update}
\end{equation}
where $v_\theta=\sigma_\theta^2$ and
$v_\phi=\sigma_\phi^2$.

A Hadamard gate exchanges the two local angular components. Since the noise is
taken symmetric, with the same strength $\sigma$ in both directions, this swap
does not change the variances and does not need to be tracked explicitly.
For asymmetric noise, the swap must be kept:
\begin{equation}
\label{SwapAxis}
(v_\theta,v_\phi)
\xrightarrow{H}
(v_\phi+\sigma_\theta^2,\;v_\theta+\sigma_\phi^2).
\end{equation}

For the rest of the single-qubit gates, only the local broadening is kept.

\subsection{Approximate Propagation of Non-Pauli Noise Through Two-Qubit Gates}
\label{ssec:propagation_two_qubit}

The CNOT gate is assumed to be noiseless for simplicity. In the exact circuit, small rotation errors produce correlated two-qubit noise after a CNOT.

For an analytical treatment, we use a simpler local rule. The noise parameters are passed through the CNOT unchanged. Each qubit keeps its local error distribution, and no new correlations are introduced. This ignores the correlated error terms generated by the entangling gate.

This approximation breaks the exact transformation of coherent errors under CNOT gates. We make this choice to keep the method simple and scalable. The goal is to estimate total circuit error rates for a given physical noise strength, not to track detailed multi-qubit correlations.

Thus, the CNOT is treated as transparent to the local Gaussian widths. The validity of this approximation is tested by comparing its predictions with the full Monte Carlo simulations already presented in Sec.~\ref{sec:CompareContPauli}.
}

\begin{algorithm}[t!]
\caption{Analytical approximation of local error propagation through
a quantum circuit.}
\label{alg:error_propagation}
\begin{algorithmic}[1]

\Require Quantum circuit with 1-qubit and 2-qubit gates in time order;
symmetric single-qubit noise strength $\sigma$

\State \textbf{State per qubit $i$:}
local variances $(v_{i,\theta},v_{i,\phi})$

\State Initialize $\forall\, i \in [1,n]:\;
(v_{i,\theta},v_{i,\phi}) \gets (0,0)$

\For{each gate in the circuit}
  \If{gate is 1-qubit gate on qubit $i$}
    \State $(v_{i,\theta},v_{i,\phi})
    \gets (v_{i,\theta}+\sigma^2,\;v_{i,\phi}+\sigma^2)$ \hfill Eq.~\eqref{eq:single_qubit_variance_update}
  \ElsIf{gate is CNOT}
    \State No local variance update
  \EndIf
\EndFor

\State \textbf{Output:}
final $(v_{i,\theta},v_{i,\phi})$ for all $i \in [1,n]$

\end{algorithmic}
\end{algorithm}

\change{\subsection{Methodology for Analyzing Noisy Quantum Circuits}
\label{ssec:analysis_methodology}

The approximate propagation model is used to estimate circuit-level error rates with reduced computational cost. It replaces gate-by-gate noise sampling with an analytic estimate of the final local error widths.

Each qubit $q_i$ is described by two small angular errors $(\theta_i,\phi_i)$, representing polar and azimuthal deviations on the Bloch sphere. For a given noise strength, the model first computes the effective widths $\sigma_{i,\theta}^{\mathrm{eff}}(\sigma)$ and $\sigma_{i,\phi}^{\mathrm{eff}}(\sigma)$ using the variance update rules in Algorithm~\ref{alg:error_propagation}.

Single-qubit gates increase the local variances by convolution with the fresh Gaussian noise. Since the noise is taken symmetric, the same variance $\sigma^2$ is added to both angular directions. Two-qubit gates do not add new noise in this approximation, and the local widths are left unchanged.

For each value $\Tilde{\sigma}$ in the noise grid, the corresponding effective widths are used to sample the final angles,
\begin{equation}
    \theta_i \sim
\mathcal{N}\!\left(0,(\sigma_{i,\theta}^{\mathrm{eff}})^2\right),
\qquad
\phi_i \sim
\mathcal{N}\!\left(0,(\sigma_{i,\phi}^{\mathrm{eff}})^2\right).
\end{equation}

One effective rotation error is then inserted at the end of the circuit on each qubit. This is repeated over $N_s$ noise samples to build the logical error curve.

This formulation keeps the variance growth of local coherent errors while avoiding explicit sampling after every faulty gate. It yields a compact circuit-level approximation that can be compared with the full noisy simulation.}

\section{Results and Discussion: Approximation}

\label{Sec:Results-AprxmModel}

In this section, we test the approximation
introduced in Sec.~\ref{sec:approximate_method}. We first apply it to the coded circuits and uncoded circuits described in Sec.~\ref{ssec:CodeCircs}. We then extend the test to random Clifford circuits introduced in Sec.~\ref{Sec:RandomCircs}. Finally, we apply the same idea to a non-Clifford example, namely the Grover search circuits in Sec.~\ref{subsec:GSACircuit}. Since Grover circuits are not Clifford circuits, we first decompose the multi-controlled \(Z\) gates into single- and two-qubit gates before applying the approximation method.

\subsection{$[[5,1,3]]$ and $[[7,1,3]]$ Circuit}

We first run the uncoded and coded circuits of Figs.~\ref{generalcircuitec=0} and~\ref{generalcircuitec=1} respectively for both codes, following the procedure of Sec.~\ref{Sec:Results-ErrorModel}. We consider a sequence of logical Hadamard circuits $m=10,50,100$, then for the circuit in Fig.~\ref{generalcircuitec=1} a round of syndrome extraction, decoding, and error correction. These simulations use the full coherent error model.

We then run the same circuits using the approximation. Instead of sampling noise after each Hadamard, we compute the accumulated noise strength analytically using Algorithm~\ref{alg:error_propagation}. It is important to mention that for the circuit in Fig.~\ref{generalcircuitec=1} the analytical approximation  is implemented until the syndrome measure circuit. There we apply a noisy syndrome measurement, as in Sec.~\ref{Sec:CodeResConPauli} since the approximate method does not take into account propagation of error from data to syndrome qubits which makes it senseless to be approximated. This means the approximation does not include the effect of error correction on the noise rate.
For each circuit instance, we sample one pair of angles per qubit at the end of the circuit. We then perform decoding and correction as before. The simulations are executed using a Monte Carlo state-vector simulator.

Each simulation produces a probability distribution over outcomes. These distributions are combined into a single distribution. The error probability $p_{\mathrm{err}}$ or logical error rate $r_{\rm logical}$ are computed using Eq.~\eqref{QEC:metric} and compared with the results from Sec.~\ref{Sec:Results-ErrorModel}.

\begin{figure}[t!]
    \centering
    \includegraphics[width=\linewidth]{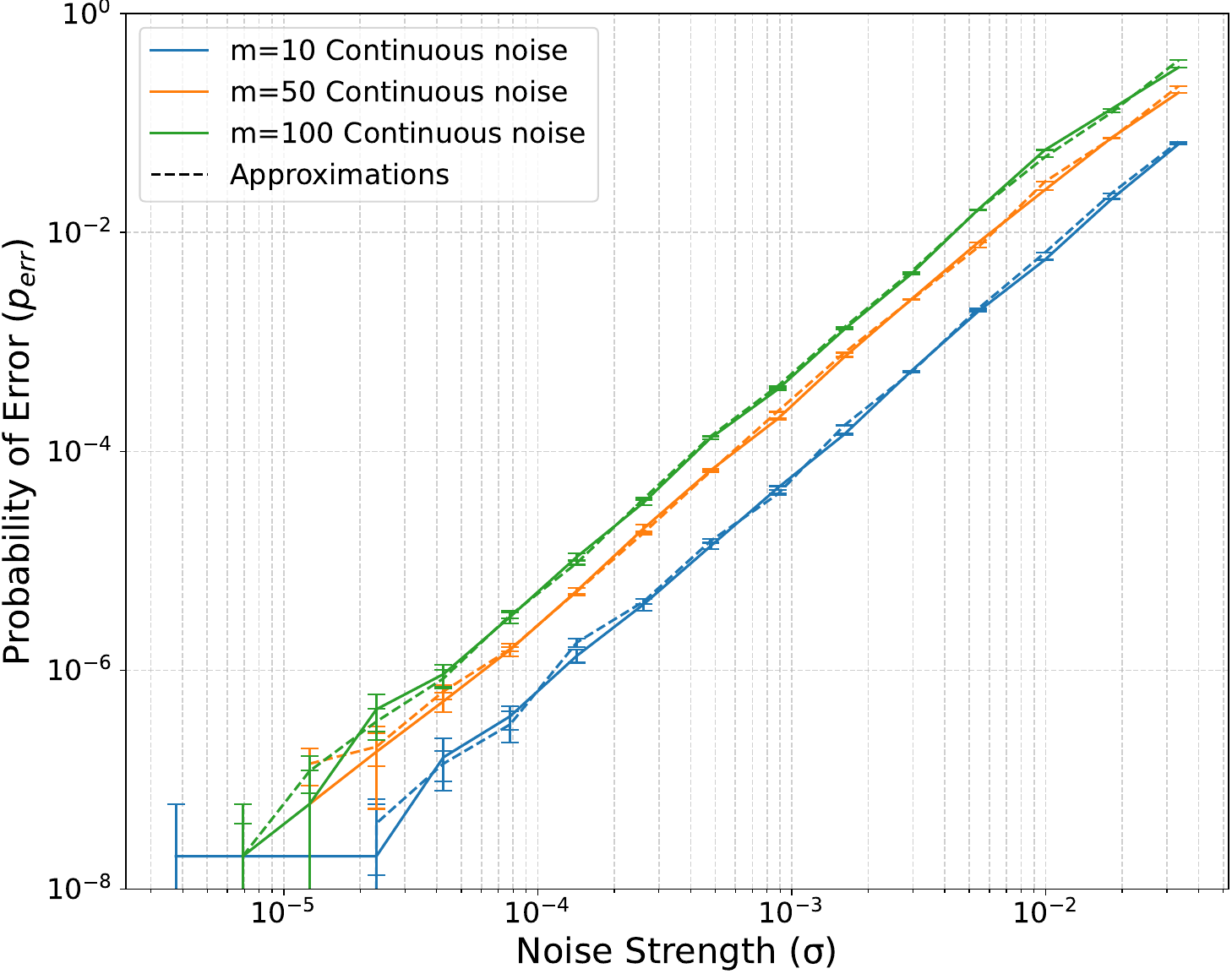}
    \caption{$[[5,1,3]]$ code: probability of error of the full simulation (solid curves) vs the error rate of the approximation simulation (dashed curves) of Fig.~\ref{generalcircuitec=0}.}
    \label{fig:resullt102}
\end{figure}

\begin{figure}[t!]
    \centering
    \includegraphics[width=\linewidth]{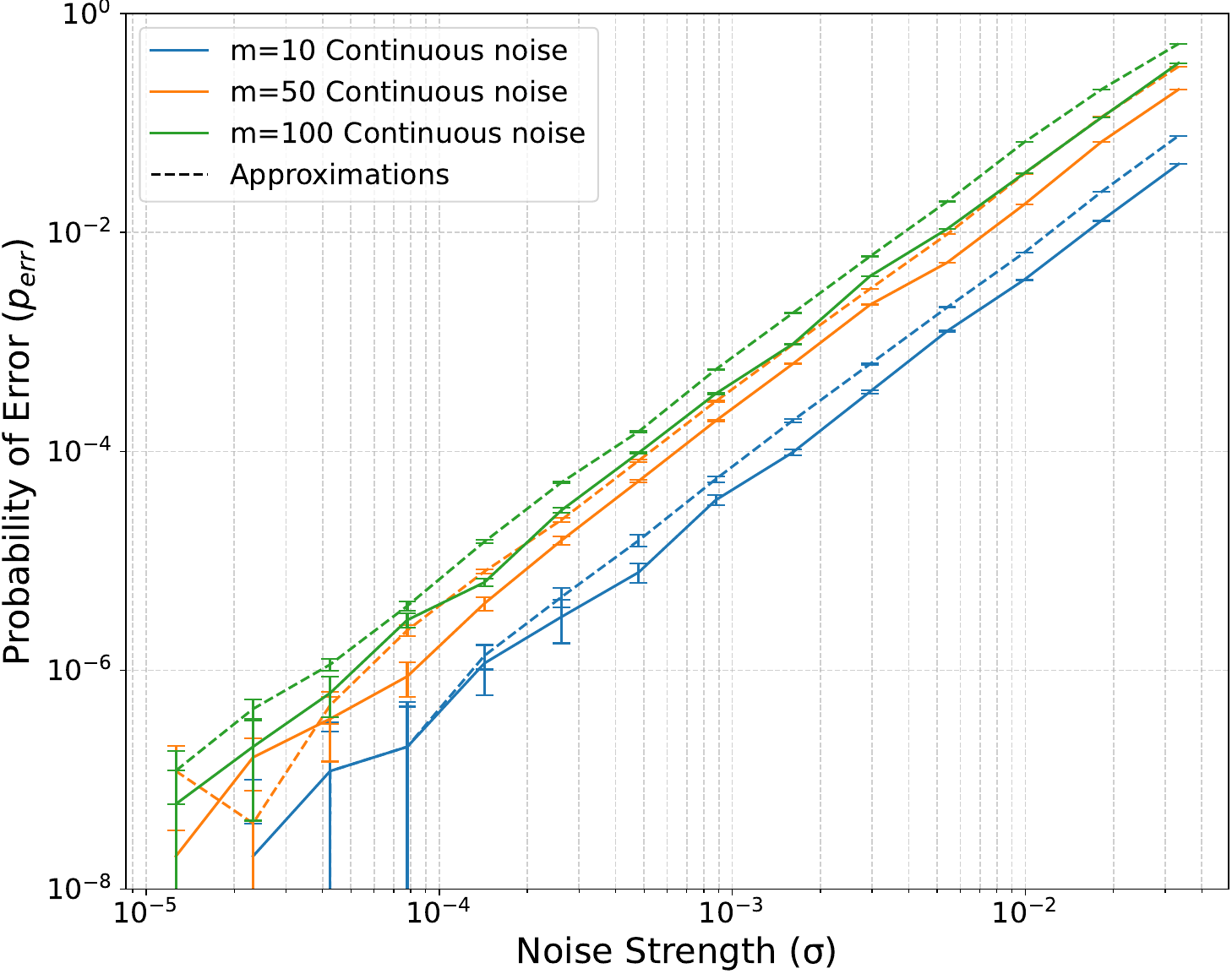}
    \caption{Same as Fig.~\ref{fig:resullt102} but for the $[[7,1,3]]$ code.}
    \label{fig:resullt103}
\end{figure}

\begin{figure}[t!]
    \centering
    \includegraphics[width=\linewidth]{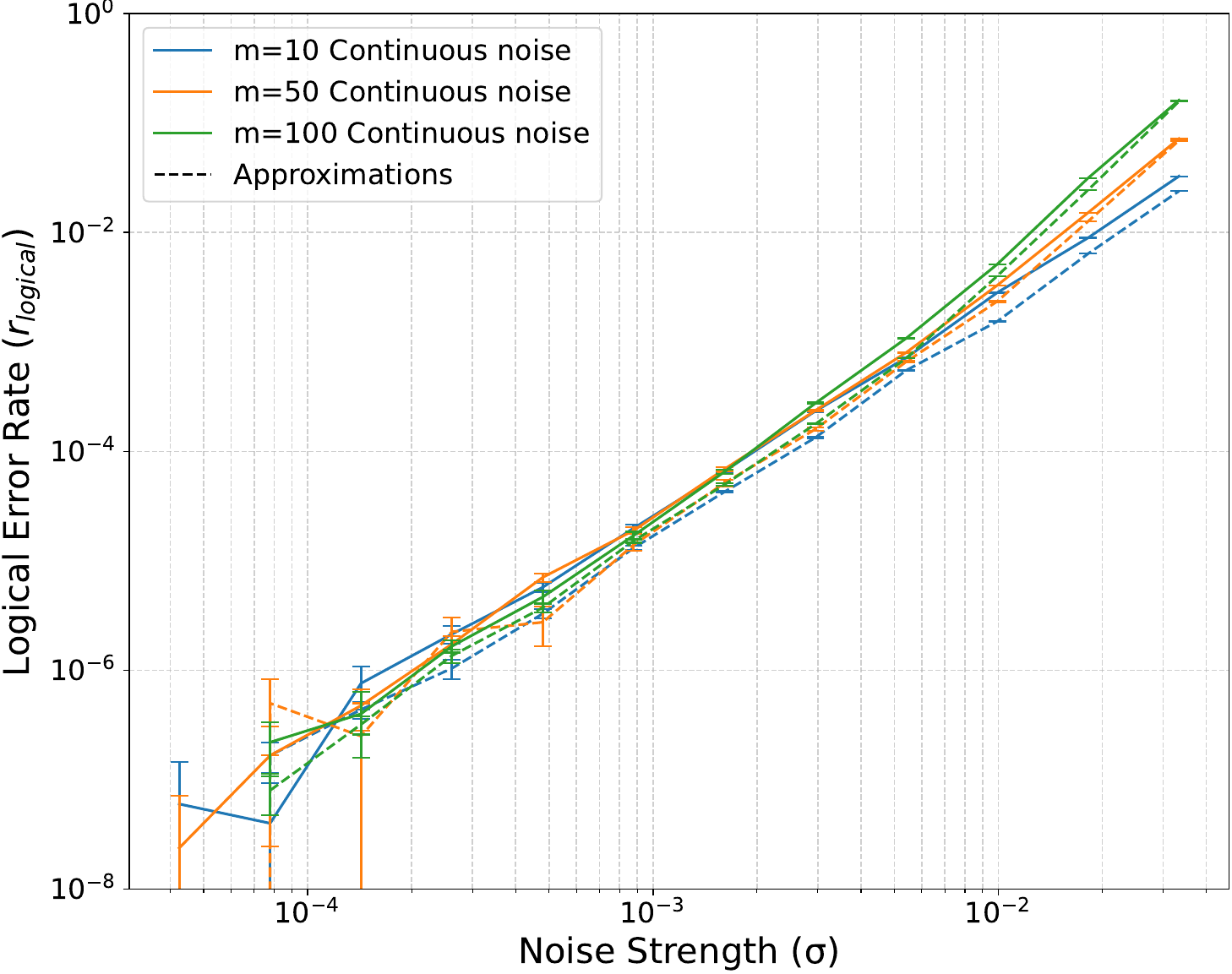}
    \caption{$[[5,1,3]]$ logical error rate $r_{\rm logical}$ of the full simulation (solid curves) vs the error rate of the approximation (dashed curves) for the circuit of Fig.~\ref{generalcircuitec=1}.}
    \label{fig:resullt202}
\end{figure}

\begin{figure}[t!]
    \centering
    \includegraphics[width=\linewidth]{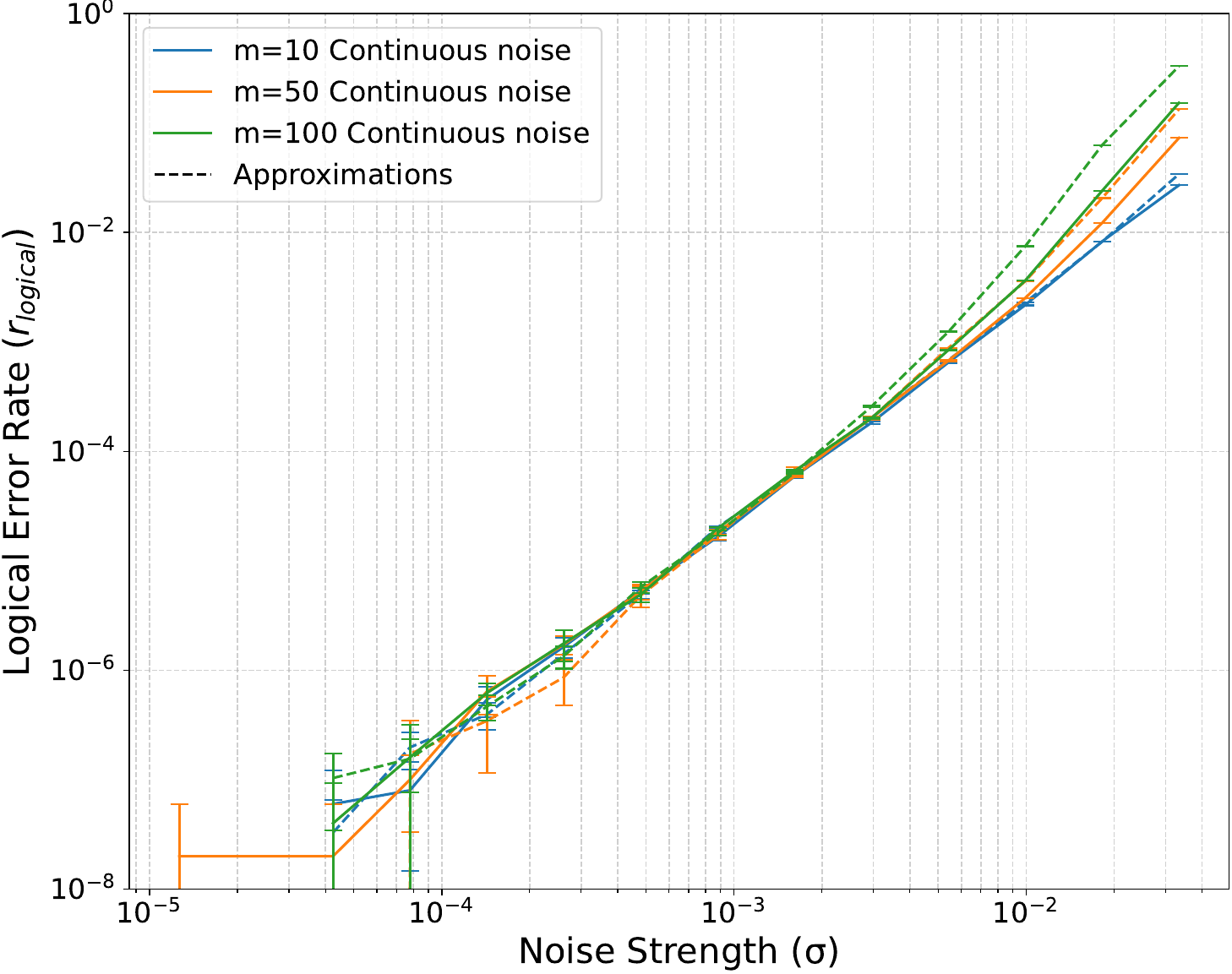}
    \caption{Same as Fig.~\ref{fig:resullt202} but for the $[[7,1,3]]$ code.}
    \label{fig:resullt203}
\end{figure}

Figures~\ref{fig:resullt102}, \ref{fig:resullt103}, \ref{fig:resullt202}, and \ref{fig:resullt203} present the results for the $[[5,1,3]]$ and $[[7,1,3]]$ codes. The solid curves correspond to the full error-model simulations, while the dashed curves of the same color show the analytical approximation.\\
Figures~\ref{fig:resullt102} and~\ref{fig:resullt103} compare circuits of different depths without syndrome extraction. The approximation follows the full error model closely over the whole range of physical error rates. For the $[[7,1,3]]$ code in Fig.~\ref{fig:resullt103}, the dashed curves lie slightly above the solid curves, indicating that the approximation tends to slightly overestimate the error probability. This behavior is not observed for the $[[5,1,3]]$ code.\\
Figures~\ref{fig:resullt202} and~\ref{fig:resullt203} correspond to the circuit shown in Fig.~\ref{generalcircuitec=1}. In these cases, the analytical approximation is applied only up to the beginning of the syndrome measurement circuit. The syndrome extraction, followed by decoding and error correction, is then simulated using the full noisy circuit. The approximation remains in close agreement with the full model over the entire range of physical error rates. The small differences between the two curves are due to the approximation itself.

In general, the results of the actual model match their approximations closely. Results for $[[7,1,3]]$ exhibit a larger deviation between the actual and approximate model than those for $[[5,1,3]]$.

The approximate treatment misses key effects of decoding and correction. It ignores correlations between qubit errors, which are essential for a complete description of error correction.



\begin{algorithm}[t!]
\caption{Run circuits with propagated Gaussian noise.}
\label{alg:approximation}
\begin{algorithmic}[1]
\Require Circuit $C$, noise grid $\Sigma=\{\Tilde{\sigma}\}$,
number of noise samples $N_s$,
\textit{Simulator} (Monte Carlo or state-vector)
\Ensure Logical error curve $\epsilon_{\mathrm{logical}}(\sigma)$
\State Compute final effective widths
  $\{(\sigma_{i,\theta}^{\mathrm{eff}}(\sigma),
  \sigma_{i,\phi}^{\mathrm{eff}}(\sigma))\}^{Nb_{qubits}}$
  using Algorithm~\ref{alg:error_propagation}
\For{$\Tilde{\sigma} \in \Sigma$}
  \State $\{(\sigma_{i,\theta}^{\mathrm{eff}},
  \sigma_{i,\phi}^{\mathrm{eff}})\}^{Nb_{qubits}} = \{(\sigma_{i,\theta}^{\mathrm{eff}}(\Tilde{\sigma}),
  \sigma_{i,\phi}^{\mathrm{eff}}(\Tilde{\sigma}))\}^{Nb_{qubits}}$
  \State Initialize \textbf{results} $\gets \varnothing$
  \For{$j = 1, \ldots, N_s$}
    \State Set $C' \gets C$
    \For{$i = 1, \ldots, Nb_{qubits}$}
      \State Sample
      $\theta_i\sim
      \mathcal{N}\!\left(0,
      (\sigma_{i,\theta}^{\mathrm{eff}})^2\right)$,
      $\phi_i\sim
      \mathcal{N}\!\left(0,
      (\sigma_{i,\phi}^{\mathrm{eff}})^2\right)$
      \Comment{Eq.~\eqref{eq:2d_gaussian}}
      \State Insert $U_{\mathrm{err}}(\theta_i,\phi_i)$
      on qubit $i$ at the end of $C'$
      \Comment{Eq.~\eqref{generalized_error}}
    \EndFor
    \State Execute $C'$ on \textit{Simulator}
    \State Append outcomes to \textbf{results}
  \EndFor
\EndFor
\State \Return \textbf{results}
\end{algorithmic}
\end{algorithm}

\subsection{Random Clifford Circuits}

\label{Sec:RandomCircRes}

Next, we test the approximation using random Clifford circuits as explained in Sec.~\ref{Sec:RandomCircs}. We first run each circuit without noise on a state-vector simulator. This produces the ideal output state, represented as a complex state vector. It serves as the reference for all comparisons.

We then simulate the same circuit with the full coherent noise model, following Algorithm~\ref{alg:coherent_noise_simulation}. After each Hadamard gate, we insert a coherent rotation error. The circuit is executed again on the same simulator.

Next, we apply the approximation model. Using Algorithm~\ref{alg:error_propagation}, we compute the final noise strength on each qubit. We sample the final error angles from the resulting distribution. These errors are added at the end of the circuit and simulated again, as described in Algorithm~\ref{alg:approximation}.

This procedure yields three output states: $\ket{\psi_{\rm ideal}}$, $\ket{\psi_{\rm model}}$, and $\ket{\psi_{\rm approx.}}$. These correspond to the noiseless circuit, the full noise model, and the approximate model. We quantify performance using the infidelity with respect to the ideal state,
\begin{gather}
\tilde{F}_{\mathrm{model}} = 1 - \left| \langle \psi_{\mathrm{model}} \mid \psi_{\mathrm{ideal}} \rangle \right|^2, \nonumber \\
\tilde{F}_{\mathrm{approx.}} = 1 - \left| \langle \psi_{\mathrm{approx.}} \mid \psi_{\mathrm{ideal}} \rangle \right|^2 .
\label{infidelities}
\end{gather}

We compare the two noise descriptions using the ratio
\begin{equation}
\label{eq:ratio}
\mathrm{ratio} = \frac{\tilde{F}_{\mathrm{approx.}}}{\tilde{F}_{\mathrm{model}}} .
\end{equation}

\begin{figure}[t!]
    \centering
    \begin{minipage}{0.48\textwidth}
        \centering
        \includegraphics[width=\textwidth]{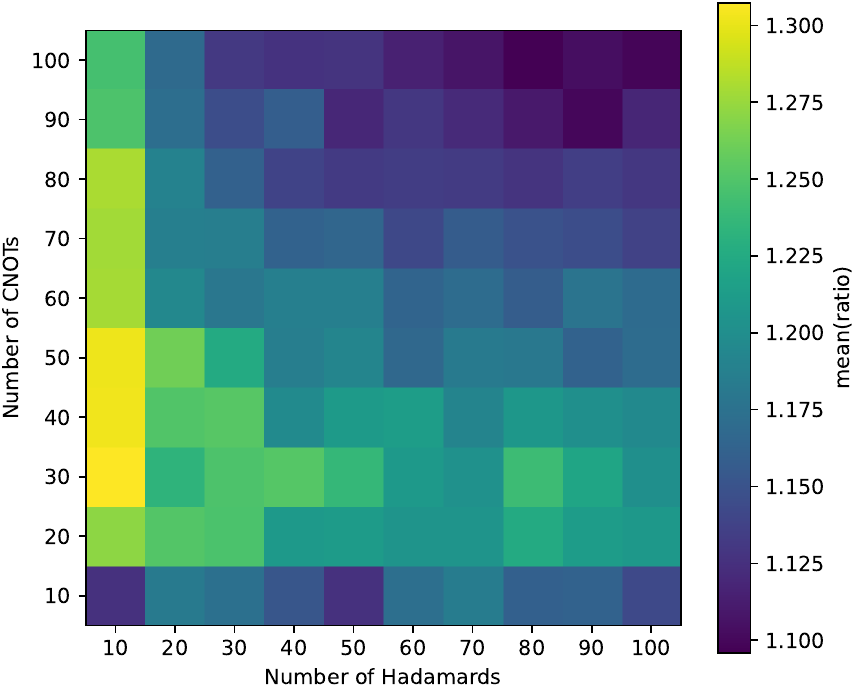}
    \end{minipage}
    \hfill
    \begin{minipage}{0.48\textwidth}
        \centering
        \includegraphics[width=\textwidth]{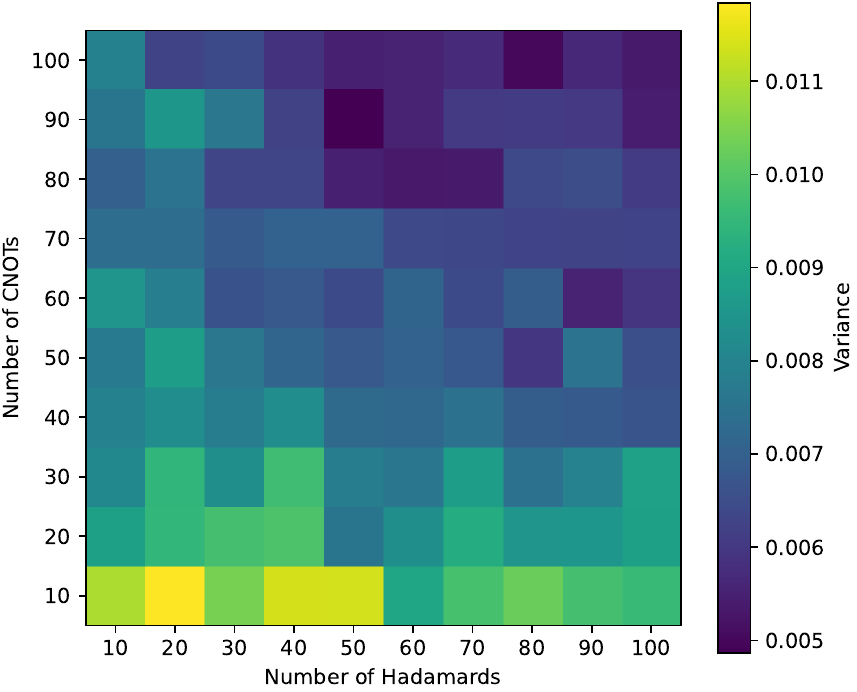}
    \end{minipage}
    \caption{Mean (top) and variance (bottom) of the ratio in Eq.~\eqref{eq:ratio} over $100$ random Clifford circuits. Each pixel corresponds to a fixed number of Hadamards and CNOTs, and both panels share the same axes.}
    \label{fig:Maps}
\end{figure}

This test is repeated over many random circuits. The number of Hadamards and CNOTs ranges from $10$ to $100$, in steps of $10$. For each pair of values, we sample $100$ random circuits. This is shown in Fig.~\ref{fig:Maps}, where each pixel represents the full and approximate simulation of $100$ different random circuits with a certain number of CNOT and Hadamard gates depending on its position in the heatmap. We compute the infidelities from Eq.~\eqref{infidelities} and the ratios from Eq.~\eqref{eq:ratio}, and finally take their mean and variance across the samples.

Fig.~\ref{fig:Maps} shows the resulting heat maps. The mean ratio remains slightly above one with a narrow variance across all depths. This indicates that the approximation accurately captures the error in random circuits. The result holds even when using fidelity-based metrics, rather than probability distributions as in Eq.~\eqref{QEC:metric}.

\subsection{Grover Search Circuits}
\label{Sec:GroverApproxRes}

We finally test the approximation on the Grover search circuits introduced in Sec.~\ref{subsec:GSACircuit}. These circuits are not Clifford circuits. The oracle and diffusion operators contain multi-controlled \(Z\) gates, so the approximation cannot be applied directly at the level of the abstract circuit. To make the circuit compatible with the gate-level error model, we first decompose each multi-controlled \(Z\) \(C^{N-1}Z\) gate into a sequence of single- and two-qubit gates.

After this decomposition, we simulate the circuit with the full coherent noise model as done in Sec.~\ref{ssec:gsaresults}. We then run the approximate model on the same decomposed circuit. The accumulated local noise strength is computed using Algorithm~\ref{alg:error_propagation}, and the final sampled errors are applied at the end of the circuit.

\begin{figure}[t!]
    \centering
    \includegraphics[width=\linewidth]{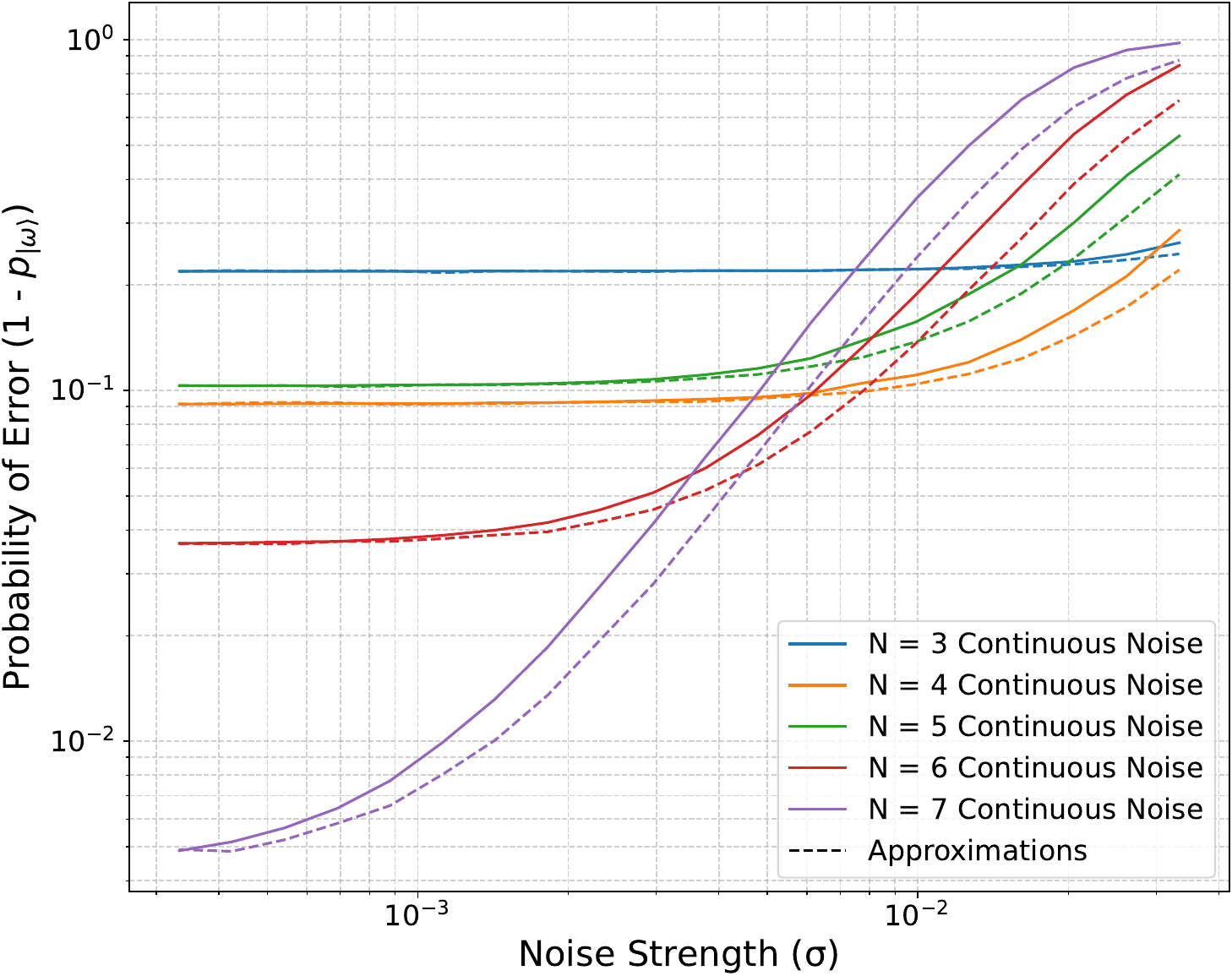}
    \caption{Error probability for Grover search circuits using the full coherent-noise model (solid curves) and the approximation model (dashed curves). The multi-controlled \(Z\) gates are decomposed into single- and two-qubit gates before applying the approximation.}
    \label{fig:GroverApprox}
\end{figure}

The error probability is computed using the Grover metric in Eq.~\eqref{GSA:metric}. We compare the full model and the approximation at the same noise strengths. The results are shown in Fig.~\ref{fig:GroverApprox}. Solid curves correspond to the full coherent-noise simulation, while dashed curves show the approximation.

The approximation follows the trend of the full model over the tested range. The main difference is a small bias: the approximation underestimates the error probability. This means that it predicts a lower noise level than the full simulation, but the deviation remains limited.

This result is useful because Grover search is not a Clifford test case. Unlike the random Clifford circuits of Sec.~\ref{Sec:RandomCircRes}, the circuit contains non-Clifford structure through the decomposed multi-controlled phase operations. The fact that the approximation still tracks the full model suggests that the method can remain accurate beyond Clifford circuits, provided the circuit is first written in the single- and two-qubit gate set used by the noise model.

\section{Conclusion}
\label{sec:conclusion}

We studied a continuous coherent noise model for quantum circuits and compared it with a discrete Pauli noise model. The coherent errors were modeled as small random rotations on the Bloch sphere, described by a von~Mises--Fisher distribution and by its Gaussian limit in the small-angle regime.

To compare both models on the same footing, we introduced a model-independent matching scheme. The continuous and Pauli channels were matched through the binary entropy at readout. This removes the direct ambiguity between the continuous noise strength and the Pauli error rate, and makes the comparison sensitive mainly to the structure of the noise.

We tested the two models on stabilizer-code circuits based on the $[[5,1,3]]$ and $[[7,1,3]]$ codes, and on Grover search circuits. In the error-corrected stabilizer-code circuits, the continuous and Pauli models yield similar logical-error trends once they are matched by entropy. This behavior is consistent with the stabilizer picture: syndrome extraction projects a general single-qubit error onto Pauli-error subspaces before recovery. Similar ideas were presented in Ref.~\cite{Beale2018}, where syndrome measurements in stabilizer codes were shown to decohere coherent errors toward effective Pauli noise. At the same time, this agreement should not be read as a general equivalence between coherent and Pauli noise. Coherent errors can produce logical effects that are not fully captured by a Pauli approximation, and this can become important after many correction cycles or at larger code distance \cite{Greenbaum2018,Barone2025}.

A similar closeness is also observed in the Grover search algorithm (GSA) tests considered in this work. Since GSA does not contain syndrome extraction, this agreement has a different origin. It suggests that, in the tested noise and depth regime, entropy matching captures the dominant effect of the noise on the success probability. The coherent structure of the continuous model does not lead to a large extra deviation from the matched Pauli model in these instances.

We also developed an approximate method for coherent-error propagation. Instead of sampling a new noisy circuit at each run, the method tracks the accumulated error distribution through the circuit. For uncoded circuits and random Clifford circuits, it agrees well with the full coherent-noise simulations while reducing the simulation cost. The random Clifford circuits act as useful stress tests, since they mix errors through different gate orders and qubit couplings. The method still tracks the main error buildup in this setting.

The approximation becomes less accurate for Grover search circuits. This is expected, since these circuits contain non-Clifford operations after the decomposition of the oracle and diffuser. The present propagation rules were built mainly for Clifford maps and can miss part of the coherent error buildup in non-Clifford gates. A more accurate treatment of these gates may therefore be needed to extend the approximation beyond Clifford-dominated circuits.

Overall, our results show that continuous coherent noise and Pauli noise can lead to similar performance trends in the regimes studied here, but this similarity is circuit-dependent.
Future work may extend the model to noisy two-qubit gates, anisotropic noise, larger codes, and different decoding strategies. Another direction is to clarify when entropy matching provides an effective link between continuous coherent noise and Pauli depolarizing noise.

The analytical approximate treatment could also be extended by using the propagated variance on each qubit to estimate the effective error probability of the whole circuit. This would avoid Monte Carlo sampling over random rotations. For error-correcting circuits, the next step is to include syndrome extraction and decoding inside the propagation model. This requires tracking which propagated error components are corrected and which ones become logical failures. An ongoing extension of this work considers incorporating syndrome extraction and decoding into the analytical approximation. Such an extension would make it possible to estimate error rates under non-Pauli noise in regimes that are not accessible to brute-force simulation.

\newpage

\begin{acknowledgments}
We would like to thank Jacopo De Nardis for fruitful discussions and for pointing out useful references.
\end{acknowledgments}

The source code and scripts required to reproduce all the results presented in this work are publicly available in the GitHub repository~\cite{github_repo}.

\bibliography{Bibliography}
\newpage

\end{document}